\documentclass[smallextended]{svjour3}      
\smartqed 

\usepackage[round]{natbib} 

\makeatletter
\def\cl@chapter{}
\makeatother
\usepackage{cleveref}
\usepackage{xspace}

\crefname{figure}{Figure}{Figures}
\Crefname{figure}{Figure}{Figures}

\newcommand{\eg}{\textit{e.g.,}\xspace}
\newcommand{\etal}{\textit{et al.}\xspace}

\usepackage{graphicx} 
\usepackage{xcolor}
\usepackage{url}
\usepackage{booktabs}
\usepackage{multirow}
\usepackage{tcolorbox}
\usepackage{subcaption}
\usepackage{enumitem}
\usepackage{threeparttable}
\usepackage{soul}
\usepackage{adjustbox}

\newif\ifDEBUG
\DEBUGfalse
\DEBUGtrue

\usepackage{soul}

\ifDEBUG
\newcommand{\JD}[1]{\textcolor{purple}{[JD:#1]}}
\newcommand{\WJ}[1]{\textcolor{olive}{[Wenxin:#1]}}
\newcommand{\CT}[1]{\textcolor{violet}{[Christoph: #1]}}
\newcommand{\HG}[1]{\textcolor{brown}{[Haoyu: #1]}}
\else
\newcommand{\JD}[1]{}
\newcommand{\WJ}[1]{}
\newcommand{\CT}[1]{}
\newcommand{\HG}[1]{}
\fi

\begin{document}

\title{AI Failures in the Eyes of the Downstream Developer: A First Look at Concerns, Practices, and Challenges}
\titlerunning{AI Failures in the Eyes of the Downstream Developer}        

\author{Haoyu Gao      \and
        Mansooreh Zahedi \and
        Wenxin Jiang \and
        Hong Yi Lin \and
        James C. Davis \and
        Christoph Treude
}
\authorrunning{Gao \etal} 

\institute{Haoyu Gao \at
              The University of Melbourne\\
              Melbourne, Victoria, Australia \\
              \email{haoyug1@student.unimelb.edu.au}          
           \and
           Mansooreh Zahedi \at
              The University of Melbourne\\
              Melbourne, Victoria, Australia \\
              \email{mansooreh.zahedi@unimelb.edu.au}  
            \and
            Wenxin Jiang \at
            Socket Inc. \\
            Wilmington, DE, USA \\
            \email{wenxin@socket.dev}
            \and
            Hong Yi Lin \at
              The University of Melbourne\\
              Melbourne, Victoria, Australia \\
              \email{tom.lin@student.unimelb.edu.au} 
            \and
            James C. Davis \at
            Purdue University \\
            West Lafayette, IN, USA \\
            \email{davisjam@purdue.edu}
            \and
            Christoph Treude \at
            Singapore Management University \\
            Singapore \\
            \email{ctreude@smu.edu.sg}
}

\date{Received: date / Accepted: date}

\maketitle

\begin{abstract}
    With the advancement of AI models, more software systems are adopting AI as a component to facilitate automation.
    Pre-trained models (PTMs) have become a cornerstone of AI-based software, allowing for rapid integration and development with lower training cost. However, their adoption also introduces failure modes such as data leakage and biased outputs, that may require careful handling by downstream developers. While previous research has proposed taxonomies of these technical concerns and various mitigation strategies, how downstream developers address these issues during the development of general AI-based software when reusing PTMs remains unexplored. Understanding downstream developers' perspectives is essential because they directly influence how these potential failures concerns translate into practice, such as determining whether immediate risks like data leakage or model bias are recognised, mitigated, or inadvertently overlooked in real-world deployments.

    This study investigates downstream developers' concerns, practices and perceived challenges regarding practical AI failures during the development of AI-based software. To achieve this, we conducted a mixed-method study, including interviews with 16 participants, a survey of 86 practitioners, and an analysis of 874 AI incidents from the AI Incident Database. Our results reveal that while developers generally demonstrate strong awareness of potential AI failures, their practices, especially during the preparation and model selection phases, are often inadequate. The lack of concrete guidelines and policies leads to significant variability in the comprehensiveness of their approaches throughout the development lifecycle, with additional challenges such as poor documentation and knowledge gaps, further impeding effective implementation. Based on our findings, we offer suggestions for AI model contributors, developers of AI-based software, researchers, and policy makers to enhance the integration of failure mitigation measures aimed at mitigating direct harms from AI failures. 
\end{abstract}

\section{Introduction}
\label{intro}
Software engineers are increasingly adopting AI models to build AI-based software systems~\citep{martinez2022software} in critical domains, including education~\citep{ciolacu2018education}, healthcare~\citep{zhou2018unet++}, and transportation~\citep{menze2015object}. With recent advancements in AI models, the expertise and resources required to train a model from scratch have increased~\citep{goyal2017accurate, wei2024jailbroken, patterson2021carbon}.
Consequently, developers are increasingly adopting pre-trained models (PTMs), which are models trained on large datasets for general tasks that can be adapted to specific applications, from platforms such as Hugging Face~\citep{jiang2023empirical, tan2022exploratory}. However, neglecting the potential failure risks of AI-based software has harmed individuals and broader societies.\footnote{\url{https://incidentdatabase.ai/}} These risks manifest as a broad range of practical AI failures, ranging from immediate technical issues such as system crashes~\citep{qi2024ai} and unreliable, hallucinated outputs~\citep{ji2023towards}, to the malicious utilisation of component vulnerabilities that trigger these crashes~\citep{gyevnar2025ai}. In this work, we focus on the specific harms caused by failures of the AI component that result in direct technical harms. Addressing these risks is an important factor to consider during the software development process. The way downstream developers handle AI failure risks throughout the development process, including data collection, fine-tuning, evaluation, deployment, and monitoring~\citep{amershi2019software}, directly shapes the risks and reliability of the final software. This raises critical questions: \textit{what risks can result from the integration of these PTMs, and how do engineers perceive and mitigate these practical AI failures?}

To address these failures, the research community has proposed taxonomies~\citep{amodei2016concrete, gao2024documenting}, evaluation methods~\citep{wang2023on}, and mitigation approaches~\citep{yiposition, roselli2019managing, ji2023towards}.  A recent meta-analysis has also compiled AI risks into a comprehensive framework, featuring both a causal taxonomy detailing how, when, and why risks occur, alongside a domain taxonomy~\citep{mitAIRisk}. Additionally, policy makers are also making efforts to provide regulations on AI risk management and responsible AI design and development~\citep{EU-ethics, isoISOIEC420012023}. While safety-critical systems often receive strict oversight, the pervasive adoption of general AI-based software means their failures can also lead to direct technical harms and real-world user consequences. Despite these efforts, we do not know how developers of general AI-based software perceive the risk of AI failures, as well as their practices and challenges during the development phase in the context of reusing PTMs. Due to the complexity of AI models and the knowledge gap between developers and model designers~\citep{nahar2022collaboration, kim2017data}, developers may approach these immediate harms differently from model designers and face distinct challenges. This is particularly true for smaller-scale company and open-source developers, where fewer resources are provided and regulations might not be enforced.  This motivates the need to examine whether and how frameworks for preventing AI failures align with or differ from the real-world development practices, as well as potential shortcomings to better inform researchers and policymakers in supporting robust failure mitigation in practice.

This study aims to investigate the concerns that downstream developers have regarding practical AI failures, the approaches they adopt, and the challenges they face throughout the entire development lifecycle of AI-based software. 
We scope our investigation to the integration of PTMs into general-purpose AI software, distinct from strictly safety-critical domains where traditional, heavyweight safety engineering practices are already mandated. To achieve this, we conducted a mixed-method study, beginning with interviews with 16 practitioners experienced in developing AI-based software, each lasting approximately one hour. We then performed a thematic analysis to derive answers to our research questions about concerns, practices, and perceived challenges when handling these potential failures. To complement the qualitative findings, we designed and conducted a survey with 86 practitioners to triangulate our results, providing quantitative evidence on practitioners’ perceived importance of different types of AI failures, the frequency of different mitigation actions, and the extent to which developers agree on various challenges in mitigating these failures. Finally, we compared developers' concerns against 874 user-reported AI failures from the AI Incident Database, identifying gaps in awareness and practices. 

Our results show that downstream developers' concerns regarding practical AI failures focus on data, model behaviour, and misuse and exploitation concerns. Additionally, their awareness of these failure risks is high in general. The practices identified by downstream developers include deeper understanding and assessment for potential failures, regular monitoring and assessment for these risks, implementing technical safeguards, maintaining documentation and expert consultation and relying on external parties and resources. However, these actions are not systematic and vary significantly in comprehensiveness. In fact, despite developers' generally high awareness of the immediate technical harms from AI failure, their actions in the early stages, particularly preparation and PTM selection, are often inadequate. The primary challenge relates to insufficient policy and regulation guidance, which is scattered across the entire development lifecycle. Other challenges, including lack of standardised benchmarks and evaluation metrics, and insufficient monitoring and maintenance methodologies, further hinder effective practices to mitigate AI failures.  
Based on our findings, we provide concrete suggestions to AI-model developers, researchers, and policy makers. For developers of AI-based software, we organise our actionable guidelines according to the different phases of the Software Development Life Cycle (SDLC). To address the knowledge transfer gap, we suggest urging AI model developers to improve documentation and by advocating for a practical failure mitigation checklist developed by researchers and policymakers, to guide downstream developers. 

In summary, the key contributions of our research are as follows.

\begin{itemize}[leftmargin=*]
    \item The first mixed-method study revealing downstream developers' concerns, practices, and challenges regarding practical AI failures emphasising the immediate technical harms.
    \item An analysis of 874 real-world incidents from the AI Incident Database through the lens of developers' concerns.
    \item Concrete suggestions for different stakeholders to improve practices to mitigate AI failures in software development.
\end{itemize}

\noindent
\textbf{\ul{Significance to Software Engineering}}:
Software is used to perform increasingly critical functions in our societies, and it increasingly incorporates AI components to do so.
Just as software engineers have reasoned about the reliability of the traditional software systems they developed, they must now reason about preventing technical failures in their AI-based systems. Currently, while the safety engineering community provides insights for critical domains, we lack the perspective of software developers building general-purpose AI systems who must translate high-level guidance into concrete mitigation practices. Our study gives a first look at their concerns, practices, and challenges, setting the stage for new knowledge and improved knowledge transfer to minimise the failure risks of the AI-based systems on which we depend. We identify the \textbf{knowledge transfer} gap affecting downstream developers. For PTM developers, improved documentation is needed to enhance transparency and usability. For policymakers, we recommend the development of evolving failure mitigation checklists, accompanied by actionable guidance linked to the most effective and up-to-date mitigation strategies. For researchers, we suggest focusing on bridging this knowledge transfer gap by investigating the specific challenges faced by downstream users, identifying the most useful types of knowledge, and exploring effective modes of knowledge dissemination.

\section{Background and Related Work}

Here we discuss software engineering practices for AI-based software (\cref{sec:background-AISoftware}). We then examine the literature on AI safety and practical AI failures (\cref{sec:background-AISafety}). Finally, we compare these concrete failures with the broader concept of AI ethics and  discuss relevant studies on developers’ perspectives to help scope and distinguish our study (\cref{sec:background-AIEthics}).

\subsection{Engineering Practices for AI-Based Software}
\label{sec:background-AISoftware}

In the ``Software 2.0'' era~\citep{karpathy2017software2}, software engineers develop \textit{AI-based software}~\citep{martinez2022software}: systems that integrate both hand-coded algorithms and learned components. The development of these learned components differs from traditional software engineering due to their data-driven nature~\citep{khomh2018software}.
\citet{nahar2022collaboration} identified interdisciplinary challenges in communication, documentation, engineering, and process when building AI-based systems.
\citet{amershi2019software} introduced a lifecycle based on experiences from Microsoft teams, highlighting three primary deviations from conventional processes: 
  (1)~ongoing discovery and management of AI models, 
  (2)~model customisation and reuse, and 
  (3)~less modular and more entangled AI components, and prone to unpredictable behaviour. The lifecycle proposed for developing AI-based software includes: model requirements, data collection, cleaning, labelling, feature engineering, model training, evaluation, deployment, and monitoring.
These differences demand new engineering practices to ensure
  robust data pipelines~\citep{rupprecht2020improving},
  model reliability~\citep{liu2023trustworthy},
  and deployment~\citep{paleyes2022challenges}.

To accelerate the development of AI-based software, engineers re-use and adapt existing \textit{pre-trained models} (PTMs)~\citep{davis2023reusing}. One major source of PTMs is the HuggingFace platform: with over 1 million models and datasets~\citep{huggingfacehub2025}, HuggingFace is the largest and least restrictive PTM registry~\citep{jiang2022ptmsupplychain}. As with other forms of software re-use~\citep{krueger1992software}, PTM re-use carries with it the challenges of selecting and evaluating a model~\citep{jiang2023empirical}, with many AI-specific details such as missing model metadata~\citep{jiang2024peatmoss}, architectural discrepancies~\citep{montes2022discrepancies}, and conversion errors~\citep{jajal2024interoperability}. Industry leaders also emphasise risks in PTM re-use and the broader AI supply chain. For example, Google extended conventional software supply chain governance to catalogue and manage AI artefacts for security, privacy, and compliance~\citep{chaudhuri2024GoogleAISSCReport}. Moreover, international governments and major public open letters from industry leaders have cautioned that rapid AI adoption may outpace the implementation of adequate safety and security controls~\citep{bletchley2023}.

These discussions underscore the challenges of developing AI-based software, whether directly or via reuse, where AI safety concerns and the immediate consequences of AI component failures have emerged as primary issues. \textit{However, literature does not describe how developers actually address these concerns.} In this work, we fill this gap by examining challenges and mitigation approaches adopted by PTM users through qualitative and quantitative analysis.

\subsection{AI Safety and AI Failures}
\label{sec:background-AISafety}

\subsubsection{Evolving Notions of AI Safety}
Any engineered system should be \textit{safe} within the parameters of its expected use~\citep{anderson1992acm}. In the traditional safety engineering community, safety has been defined strictly in terms of preventing unacceptable risks of death, injury, loss of property, or environmental harm~\citep{leveson1983analyzing}. According to the IEC 61508 standard, safety is defined as the freedom from unacceptable risk of physical injury or damage to the health of people, property, or the environment~\citep{iec61508}. In this classical context, \textit{functional safety} specifically ensures that a system operates correctly in response to its inputs to mitigate these severe physical risks. Therefore, within this traditional framework, safety is treated as a distinct system property focused exclusively on preventing catastrophic physical and environmental losses, establishing a firm boundary for what constitutes a safety failure.

However, as AI models are increasingly integrated into general-purpose software, the terminology surrounding ``safety'' has fractured. Some recent literature attempts to maintain strict definitional boundaries; for instance, \citet{qi2024ai} argue that AI risk management should formally separate safety, which prevents the system from harming its environment, from security, which protects the system against malicious external actors. However, other frameworks advocate for a broader, pluralistic understanding of AI safety that captures a wide array of technical harms and incorporates traditional security concerns~\citep{gyevnar2025ai}. There is also literature that refers to AI safety in terms of existential risks, such as societal collapse~\citep{lazar2023ai}. Within this expanded discourse, a central theme has emerged: the focus on practical AI failures that cause direct harm, rather than indirect threats such as existential risks. As \citet{yampolskiy2016artificial} argue, an AI safety mechanism cannot be considered functionally safe if it is not designed to resist attacks by malevolent human actors, demonstrating the necessity of evaluating both accidental safety flaws and intentional security vulnerabilities simultaneously as instances of AI failure. This broadened perspective is also increasingly formalised in emerging industry guidelines. For example, the UL 3115~\citep{ul3115} Outline of Investigation for Safety of AI-Based Products explicitly applies its safety evaluation framework to both physical ``AI-embedded products'' (\eg autonomous driving and medical devices) and entirely "digital AI products" (\eg LLM-based customer service applications). This industry shift acknowledges that technical vulnerabilities and biases in purely digital spaces now constitute critical product safety concerns.

Reflecting this broader, unified perspective, researchers have described many practical aspects of AI failure that are now commonly categorised under AI safety. These encompass unintended harms, such as discrimination~\citep{sheng2021societal}, misinformation~\citep{zhang2023siren}, and over-reliance or unsafe use of software~\citep{weidinger2021ethical}, alongside non-accidental vulnerabilities stemming from malicious use or adversarial attacks~\citep{hernandez2020ai}. With the rapid advancement and increasing real-world deployment of AI technologies, concrete AI failures are emerging in the delivered software products. A recent systematic literature review on AI safety underscores the gaps in human value alignment, ethical governance, and safety frameworks for AI system implementation~\citep{salhab2024systematic}. Additionally, production-grade chatbots such as ChatGPT and Gemini have raised concerns in potential failures, including policy compliance issues~\citep{hacker2023regulating}, susceptibility to generating or endorsing violent content~\citep{zhuo2023red}, and systemic bias~\citep{fleisig-etal-2024-linguistic}. 

To systematically track and understand these emerging issues, the research community has begun cataloguing real-world AI failures and mapping risk landscapes. For instance, the AI Incident Database (AIID) serves as a collective memory of harms or near-harms caused by deployed AI systems~\citep{mcgregor2021preventing}. Parallel to this incident tracking, the MIT AI Risk Repository synthesises various risk frameworks into a structured taxonomy of causes and domains~\citep{mitAIRisk}. While these repositories provide empirical evidence of failures and categorise risks, they do not capture how downstream software developers actively perceive and prevent these issues during their daily engineering workflows. To navigate this terminological ambiguity, our study focuses specifically on practical AI failures~\citep{yampolskiy2016artificial}.

\begin{tcolorbox}[left=1pt, top=1pt, right=1pt, bottom=1pt] \textbf{AI Failure Definition}: 
Drawing on the AI Incident Database~\citep{mcgregor2021preventing}, we operationalise AI failures as instances where an AI component in the system produces unintended, incorrect, or unsafe behaviour that results in immediate technical harms. This encompasses a broad spectrum of practical issues, ranging from accidental AI component mistakes and unpredictable environmental interactions to vulnerabilities deliberately exploited by malicious actors.
\end{tcolorbox}

\subsubsection{Mitigation Frameworks}

 Beyond theoretical definitions, mitigating these issues in practice requires concrete mitigation strategies. In the context of traditional software engineering, system reliability and failure prevention have been a longstanding goal~\citep{lutz2000safetyroadmap} promoted through controlled engineering processes and standards such as IEC 61508~\citep{iec61508}, which defines Safety Integrity Levels (SILs) to measure confidence in the safe operation of software-intensive systems. The safety engineering community has further developed structured methodologies for reasoning about program hazards, such as fault tree analysis~\citep{leveson1983analyzing} and system-theoretic approaches~\citep{leveson2004new}, as well as safety case frameworks that provide structured arguments for why a system is acceptably safe~\citep{kelly1999arguing}. Researchers have also proposed safety-aware techniques across the entire engineering cycle, \eg during requirements engineering~\citep{martins2016requirements}, design~\citep{leveson1983analyzing}, implementation~\citep{chong2021code}, validation~\citep{gladisch2019experience}, and failure analysis~\citep{anandayuvaraj2024fail}.

Transitioning to the mitigation of AI-specific technical failures, researchers have proposed novel methods for approaching these concerns during the development of AI-based software. These include dataset construction methods~\citep{wang2024beancounter}, evaluation of AI safety risks~\citep{wang2023on}, mitigation strategies~\citep{jung2024a, chen2023holistic}, and monitoring techniques~\citep{naveed2024towards}. Additionally, comprehensive governance frameworks have been proposed to address reliability of AI components and defect prevention at a systemic level. For instance, \citet{agarwal2024sevenLayerModelforAIFairnessStandard} introduced a seven-layer model to standardise fairness assessment, and~\citet{lu2024responsibleAIPattern} described responsible AI patterns for governance and engineering. Others have surveyed bias in AI systems across scientific disciplines~\citep{mcgovern2024identifyingAIBiasforEarthScience} and proposed governance structures such as independent audits~\citep{falco2021governingAISafetyThroughIndependentAudits} and ethics-to-practice frameworks~\citep{shneiderman2020bridgingGapBetweenEthicsandPractices, amodei2016concrete}. Furthermore, policies such as the EU AI Safety Standards~\citep{pouget2024EUAISafetyStandards} aim to regulate high-risk AI systems. On an international level, ISO/IEC 42001~\citep{isoISOIEC420012023} offers a standard framework for organisations to address risks related to transparency, robustness, and accountability. However, such policies and frameworks often remain high-level and do not provide concrete action suggestions for practitioners to mitigate failures in everyday development~\citep{pouget2024EUAISafetyStandards}. To help address this gap between high-level regulatory principles and practical engineering implementation, recent industry efforts, such as the emerging UL 3115 standard~\citep{ul3115}, aim to offer a more structured evaluation framework. Rather than acting as a definitive solution, it outlines a cross-domain set of safety guidelines—touching upon concrete aspects like model robustness, data privacy, bias identification, and transparency—that are intended to be integrated with existing domain-specific end-product standards.

\subsubsection{Gap Analysis}
\textit{Despite ongoing efforts from AI researchers and policymakers to address these technical risks, little attention has been paid to understanding the current engineering practices of downstream developers on approaching these failure risks}. Gaining insights from their perspective is crucial, as it reveals the state of practice and the challenges in addressing practical AI failures. This understanding helps identify and bridge gaps in knowledge (and knowledge transfer) between AI research, regulations, and real-world software engineering practices.

\subsection{Developers' Perspectives on Software Failures}\label{sec:background-AIEthics}

Regarding traditional software failures, a wealth of literature explores structured developer practices in debugging, incident response, and post-mortem analyses, highlighting how established safety frameworks~\citep{sillito2020failures,anandayuvaraj2025learning} and past failure knowledge actively advocate developers' mitigation efforts~\citep{anandayuvaraj2023incorporating}. However, as software systems increasingly incorporate pre-trained models, the nature of software failures expands to include AI-specific vulnerabilities.

Currently, there is a lack of empirical research exploring how software developers view and mitigate concrete AI failures. The closest existing literature examines developer perspectives through the broader lens of AI ethics. While the practical AI failures we investigate sometimes contain ethical implications, they are distinct from this broader discourse in their level of abstraction. AI ethics is broad in scope, defined by Siau \& Wang~\citep{siau2020artificial} as ``the principles of developing AI to interact with other AIs and humans ethically and function ethically in society''. Governmental regulations, such as those in Australia~\citep{AUS-ethics} and the EU~\citep{EU-ethics}, often frame AI ethics through this macro-perspective, encompassing human rights, social well-being, and overarching corporate accountability. In contrast, while our adopted definition of AI failures (Section~\ref{sec:background-AISafety}) includes practical manifestations of ethical issues, such as algorithmic bias or toxic outputs, it approaches them strictly through an engineering and risk-mitigation lens. AI failures focus on the concrete technical vulnerabilities and system failures that developers must actively prevent, rather than the abstract moral philosophy or organisational values surrounding the technology.

Consequently, while previous empirical studies have explored developer perspectives within the broader scope of AI ethics, they often focus on high-level moral or organisational challenges. For example, \citet{pant2024ethics} conducted a closed-ended survey with 100 general AI practitioners to examine their awareness of AI ethics and the challenges they face, identifying general, technology-related, and human-related hurdles. \citet{ali2023walking} interviewed 25 AI ethics entrepreneurs to investigate the practical difficulties of implementing AI ethics within technology companies. Their findings highlight systemic organisational issues, such as unstable team structures, the decoupling of practices from ethical principles, and the low prioritisation of ethical concerns. Similarly, \citet{wong2021tactics} investigated how user experience (UX) professionals attempt to reshape the values embodied and promoted by their companies. Different from the previously mentioned works on broad ethical frameworks, our study focuses specifically on practical AI failures, operationalised through the lens of direct, technical harms rather than indirect or speculative harms such as existential risks. Rather than targeting general AI practitioners, we investigate the dominant pattern of PTM reuse in the development of AI-based software~\citep{banyongrakkul2025release}. Accordingly, we recruited participants who help embed intelligent components into the systems they ship, where those components were derived in some way from PTMs. This study offers a focused perspective on downstream developers engaged in PTM reuse for the development of AI-based software, particularly in the context of the entire SDLC.

\section{Research Questions \& Methodology}\label{sec:methodology}

We structured our research into three research questions (RQs).

\begin{figure}[h]
    \centering
    \includegraphics[width=\columnwidth]{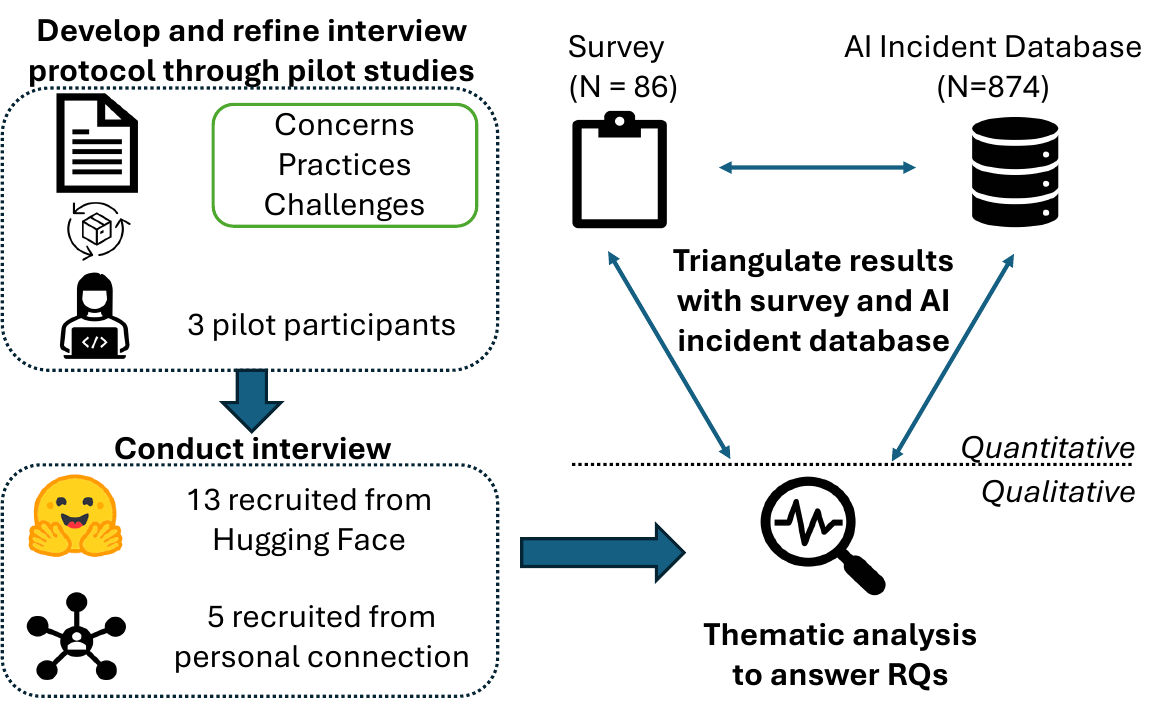}
    \caption{
    Study design and methodology
    }
    \label{fig:method-process}
\end{figure}

\begin{itemize} [leftmargin=30pt, rightmargin=5pt]
\item [\textbf{RQ1:}] What are developers' concerns regarding AI failures in the development of AI-based software?
\item [\textbf{RQ2:}] How do developers approach and mitigate these potential AI failures during the development of AI-based software?
\item [\textbf{RQ3:}] What challenges do developers perceive when handling AI failures as they develop AI-based software?
\end{itemize}

These three research questions are designed to capture a comprehensive view of how developers manage technical risks throughout the SDLC. We formulated RQ1 to establish the empirical baseline of the problem space, aiming to understand the specific, practical failures that practitioners actually anticipate and encounter in the wild. Once these focal points are identified, RQ2 investigates the current state of practice to reveal the concrete mitigation strategies developers actively employ. Finally, RQ3 explores the systemic hurdles and technical roadblocks that hinder effective mitigation, highlighting the critical gaps where future research, tooling, and policy interventions are most urgently needed.

\textbf{Study Design}. This study is structured around the definition in~\ref{sec:background-AISafety}. Our instruments and analyses use this definition in various ways, which we elaborate in each subsection.
To address these RQs, we conducted a mixed-method study as illustrated in Figure~\ref{fig:method-process}. Mixed-method research is appropriate for our context, as downstream developers’ perceptions and practices involve both human and technical dimensions that cannot be fully captured by a single qualitative or quantitative method; instead, the two perspectives offer complementary strengths~\citep{easterbrook2008selecting}.  We began by interviewing 16 participants, as interviews are well-suited for exploring complex experiences and eliciting rich, detailed insights from downstream developers~\citep{seaman1999qualitative}. This was followed by a survey with 86 participants to validate and quantify the findings~\citep{aniche2018modern}, and an analysis of 874 incident cases from the AI Incident Database to triangulate our results against real-world AI failures~\citep{anandayuvaraj2024fail}.

Our study was conducted with oversight by our institution's ethics board. Due to its constraints, we cannot share the raw interview transcripts. However, 
all study instruments, detailed interview analysis, raw survey responses, and all data about AI incident analysis are available in our artefact (\cref{sec:replication}).

\subsection{Method 1: Interviews}\label{sec:interview-method}

\textbf{Instrument Design.} We designed a semi-structured interview protocol exploring the three RQs. Following the development process of AI-based software outlined by \citet{amershi2019software}, we structured the interview into sections covering background understanding, model selection, model development and testing, model deployment and maintenance. For each stage, we prepared questions asking  participants' development experience, concerns, practices, and challenges, as indicated in Table~\ref{tab:sample-questions}. With the semi-structured interview format, at each stage we first explored the participants' past development experience, followed by their concerns regarding practical AI failures, the practices they adopted to mitigate these direct technical harms associated with the failures, and the related challenges encountered during the process. This approach situates their responses within specific development contexts, reducing the risk of concept misinterpretation during analysis.

\begin{table}[t]
\centering
\caption{
Examples of participants operationalising the concept of AI safety into concrete technical failures. These emerged from the broad, open-ended interview prompt: ``What are your considerations regarding AI safety, and what are your perspectives on it?''}
 \resizebox{\columnwidth}{!}{%
\begin{tabular}{p{0.08\linewidth} p{0.85\linewidth}} \toprule \textbf{ID} & \textbf{Participant Quote (Operationalising AI Safety'')} \\
\midrule

\textbf{P4} & \textit{``One project that I participate in is to seek if the response give the safe answer, not like the answers that are against the local laws, or be very offensive. So this is the safety aspect that we focus on.''} \\
\midrule
\textbf{P14} & \textit{``I believe the most important safety thing to have in mind... is when we integrate some model with a pipeline that involve... an external user... there are a lot of vulnerabilities that can be seen, like jailbreaking your model or trying to steal some data... [other concerns include] copying and pasting code from an LLM output... and put a really critical bug into the system... racist [outputs]... Hallucination as well... That four things that came to my mind right now are the most critical safety concerns we have faced.''} \\
        \midrule
\textbf{P18} & \textit{``When I think of safety, personally I think of things like privacy... whether or not the responses are grounded or if they're supported by documents, things like hallucination... My work mainly focuses around privacy and... hallucination [and] factuality type.''} \\

\bottomrule
\end{tabular}}
\label{tab:quote-examples}
\end{table}

Our interview protocol used the broader term ``AI safety'' to prompt discussion about risks and concerns in AI-enabled software systems, because it remains a prevailing vocabulary among industry practitioners~\citep{nistRiskManagement}. During the interviews, we observed how participants operationalised this broad term within their own engineering contexts. Rather than engaging with the abstract ``AI safety'' discourse, practitioners consistently interpreted the concept as the need to mitigate concrete, local engineering failures, such as hallucinations, privacy leaks, systemic bias, and reliability problems. Consequently, our RQs and analysis use the term ``AI failure'' to follow the participants' lead. Instead of assuming a shared, top-down definition of AI safety, we focused our thematic analysis strictly on the practical failures and technical risks they actively described. To illustrate this organic operationalisation, Table~\ref{tab:quote-examples} provides representative examples of how participants translated the broad prompt of ``AI safety'' into specific, local engineering failures at the beginning of our interview. The comprehensive mapping of these operationalisations for all 16 analysed participants is available in our replication package (\cref{sec:replication}).

If needed, during the interview, we steered the conversation further into the direction of these concrete technical failures. For example, when the participants' answers digressed into indirect harms, such as carbon emissions or super-intelligence replacing humans (``singularity''), we calibrated the response. Subsequently, because our findings were empirically grounded in the participants' own technical realities, providing a theoretical definition in the survey phase was unnecessary. Instead of interpreting the abstract ``AI safety'' concept themselves, survey participants simply evaluated concrete, summarised AI failure scenarios derived directly from the interview themes.

We conducted three pilot interviews with developers of AI-based software from our industrial connections, and then applied minor tweaks to our interview protocol based on the results. The final version of our interview took approximately one hour. Table~\ref{tab:sample-questions} shows sample questions.

\begin{table}[]
    \centering
    \caption{
    Summary of interview and survey instruments. (See complete data in replication package in~\cref{sec:replication}). (The ``AI failure concerns'' in square brackets are replaced with the specific AI failure scenario in the interview)}
    
    \resizebox{\columnwidth}{!}{%
    \begin{tabular}{l l l}
    \toprule
        \textbf{Topic (\# Questions)} & \textbf{Sample Questions} & \textbf{Survey Scales}  \\
        \midrule
        Background (5) & What is your role in your team? & N/A\\
        \midrule
        Concerns (4) & \begin{tabular}[c]{@{}l@{}}What concerns regarding poten-\\ tial AI failure issues would you \\ consider at model selection stage?\end{tabular}  &\begin{tabular}[c]{@{}l@{}}5-point scale: From ``Not\\ important at all''  to\\  ``Extremely important''\end{tabular}\\
        \midrule
        Practices (12) & \begin{tabular}[c]{@{}l@{}}Identifying the [AI failure concern] \\ in the \textit{deployment} phase, what \\ would you do next to approach it? \end{tabular} & \begin{tabular}[c]{@{}l@{}}4-point scale: From ``I \\ don't do this at all''\\ to ``I prioritise this \\ as a critical task''\end{tabular} \\
        \midrule
        Challenges (3) & \begin{tabular}[c]{@{}l@{}}What challenges do you face\\ when trying to consider and\\  handle [AI failure concerns] \\ in the \textit{development} phase?\end{tabular} & \begin{tabular}[c]{@{}l@{}}5-point scale: From \\ ``Strongly disagree'' to \\ ``Strongly agree'' \end{tabular}   \\
        \bottomrule
    \end{tabular}}
    
    \label{tab:sample-questions}
\end{table}

\textbf{Recruitment.} To access downstream developers actively engaged in the development of AI-based software, we recruited participants from the Hugging Face ecosystem and our professional network, targeting those with development experience in AI-based software. Hugging Face is the largest registry for PTMs; therefore, its ecosystem provides direct access to developers actively engaging in PTM reuse, which aligns perfectly with the scope of our study. Following the design of a previous study to recruit participants from Hugging Face~\citep{jiang2023CVReengineering}, we extracted all PRO users and members of organisations with over 50 users, retaining only those labelled as \textit{company} or \textit{non-profit}, excluding labels such as \textit{community} and \textit{university}.
Since Hugging Face profiles sometimes link to personal websites, we contacted users via publicly available emails on their personal websites to comply with
site policies~\citep{githubGitHubAcceptable}. This process yielded 98 PRO users and 783 organisational users, totalling 868 unique users after deduplication.

We first contacted PRO users, followed by organisational users, prioritising  from larger organisations based on the number of members. Recruitment and analysis occurred concurrently.
In total, we contacted 667 users on Hugging Face.
We scheduled interviews with 17 users, of whom 13 attended.\footnote{We did not specifically target participants developing ``safety-critical'' systems (\eg medical devices~\citet{knight2002safety}) during the recruitment process, because practical AI failures (\cref{sec:background-AISafety}) encompasses a broad spectrum of technical harms in general-purpose software than physical safety.} 
This resulted in a response rate of 2.55\% similar to other software engineering studies~\citep{lethbridge2005studying, singer2014software}. During the interviews, 2 participants (P3 and P5) did not provide useful responses, and we discarded them from the analysis. To complement this sample, we invited five industry developers of AI-based software from our personal network, totalling 18 participants, with 16 being analysed. 
Table~\ref{tab:demographics} shows the demographics of our interview participants, in which we keep the ID for all 18 participants, but removed P3 and P5 from the analysis.
Each participant was compensated \$20 USD for their time. Additionally, interviews were conducted and analysed by a researcher with technical and qualitative training, including experience in AI-based software development and AI failure analysis. This background is essential in conducting semi-structure interviews, as it enables us to ask effective follow-up questions.

\begin{table}[]
    \caption{
    Interview participant demographics.
    We measured experience in Deep Learning (DL) and Software Engineering (SE) in years. See~\cref{sec:data-characteristics} for safety-critical definition.
    P3 and P5 were excluded from the analysis because they did not provide useful responses.
    }
    \centering
\resizebox{\columnwidth}{!}{%
    \begin{tabular}{r r r r r r r r}
    \toprule
        \textbf{ID} & \textbf{DL Exp.} & \textbf{SE Exp.} & \textbf{\# AI software} & \textbf{Degree} & \textbf{Org. Size} & \textbf{Domain} & \textbf{Safety-Critical}  \\
        \midrule
         P1 & 1 & 15 & 1-5 & Bachelor & Open-source & NLP & No\\ 
         P2 & 6 & 6 & 5-10 & PhD & Small & CV & No\\
         P3 & 2 & 4 & 1-5 & Bachelor & Open-source & NLP & No \\
         P4 & 1 & 2 & 1-5 & PhD & Large & NLP & No\\
         P5 & 2 & 7 & 1-5 & PhD & Open-source & CV & No\\
         P6 & 12 & 25 & 1-5 & PhD & Small & NLP & No\\
         P7 & 4 & 2 & 10-25 & PhD & Large & NLP & Yes\\
         P8 & 2 & 2 & 25+ & Master & Large & CV & No\\
         P9 & 6 & 6 & 1-5 & PhD & Large & CV, NLP & No\\
         P10 & 7 & 4 & 5-10 & PhD & Large & CV & No\\
         P11 & 4 & 3 & 25+ & Master & Medium & CV, NLP & No\\
         P12 & 4 & 4 & 25+ & Master & Large & CV, NLP & No\\
         P13 & 3 & 2 & 5-10 & Bachelor & Large & NLP & No\\
         P14 & 1 & 2 & 5-10 & PhD & Small & NLP & Yes\\
         P15 & 3 & 3 & 1-5 & Bachelor & Medium & NLP & No\\
         P16 & 1 & 2 & 25+ & Master & Small & NLP & No\\
         P17 & 4 & 4 & 25+ & PhD & Small & NLP & No\\
         P18 & 7 & 7 & 10-25 & PhD & Medium &  NLP & No\\
        \bottomrule

    \end{tabular}}
    
    \label{tab:demographics}
\end{table}

\textbf{Analysis.} Following suggestions in software engineering interview studies~\citep{hove2005experiences}, the first two sessions were conducted by the first author and a senior researcher, who is also co-author of this study. They discussed the interview flow, after which the first author conducted the remaining sessions. We transcribed the interview recordings using Microsoft 365's transcription tool, verified manually, and organised in NVivo for analysis. 

Regarding the analysis procedure, we adopted suggested steps from thematic analysis~\citep{braun2006using, cruzes2011recommended} for the qualitative analysis stage of the interview transcripts. Two authors, including the first author and the author who participated in two interview sessions and closely monitored the interview transcripts, first read through two interview transcripts, summarising the content as key points related to practical AI failures, practices, and challenges that corresponded to the three RQs, as suggested to initiate the open coding~\citep{hoda2012self}. They then proceeded to assign base level of codes to the extracted key points. Subsequently, we developed higher level abstractions of subthemes by grouping relevant codes together and then the highest level abstractions of themes. Consensus was reached before the first author continued to apply this process to the rest of the transcripts.  During the coding process, we constantly compared the emerging codes within and across interview transcript key points, with regular discussions held to verify the code assignment. This iterative process of code development led to continuous adjustment of the codes. Grouping of the codes into themes was conducted concurrently, with each subtheme and theme discussed and adjusted as well.  

\begin{figure}
    \centering
    \includegraphics[width=\columnwidth]{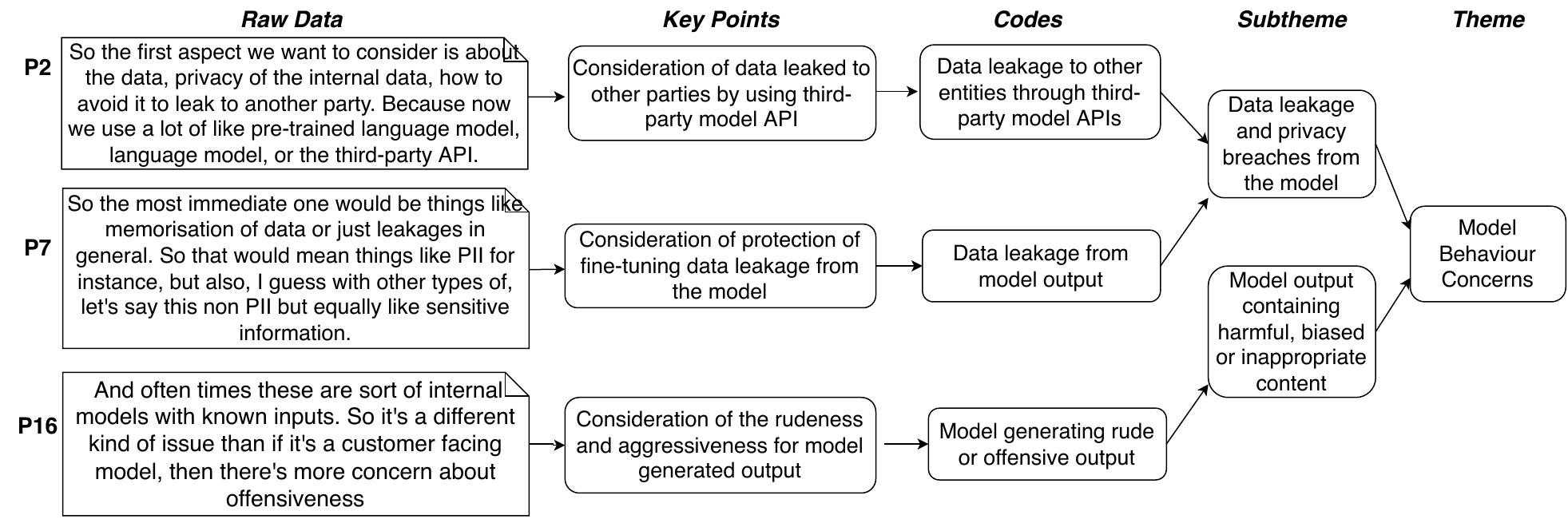}
    \caption{
    Illustration of our coding process
    }
    \label{fig:coding-process}
\end{figure}

We derived 29, 100, and 55 codes for the three RQs, respectively, based on the above procedure. Three authors independently reviewed the coding process and the synthesised subthemes and themes for each research question. The results received agreement among the entire group. Figure~\ref{fig:coding-process} displays a sample of the coding process.

\begin{figure}
    \centering
    \includegraphics[width=0.75\linewidth]{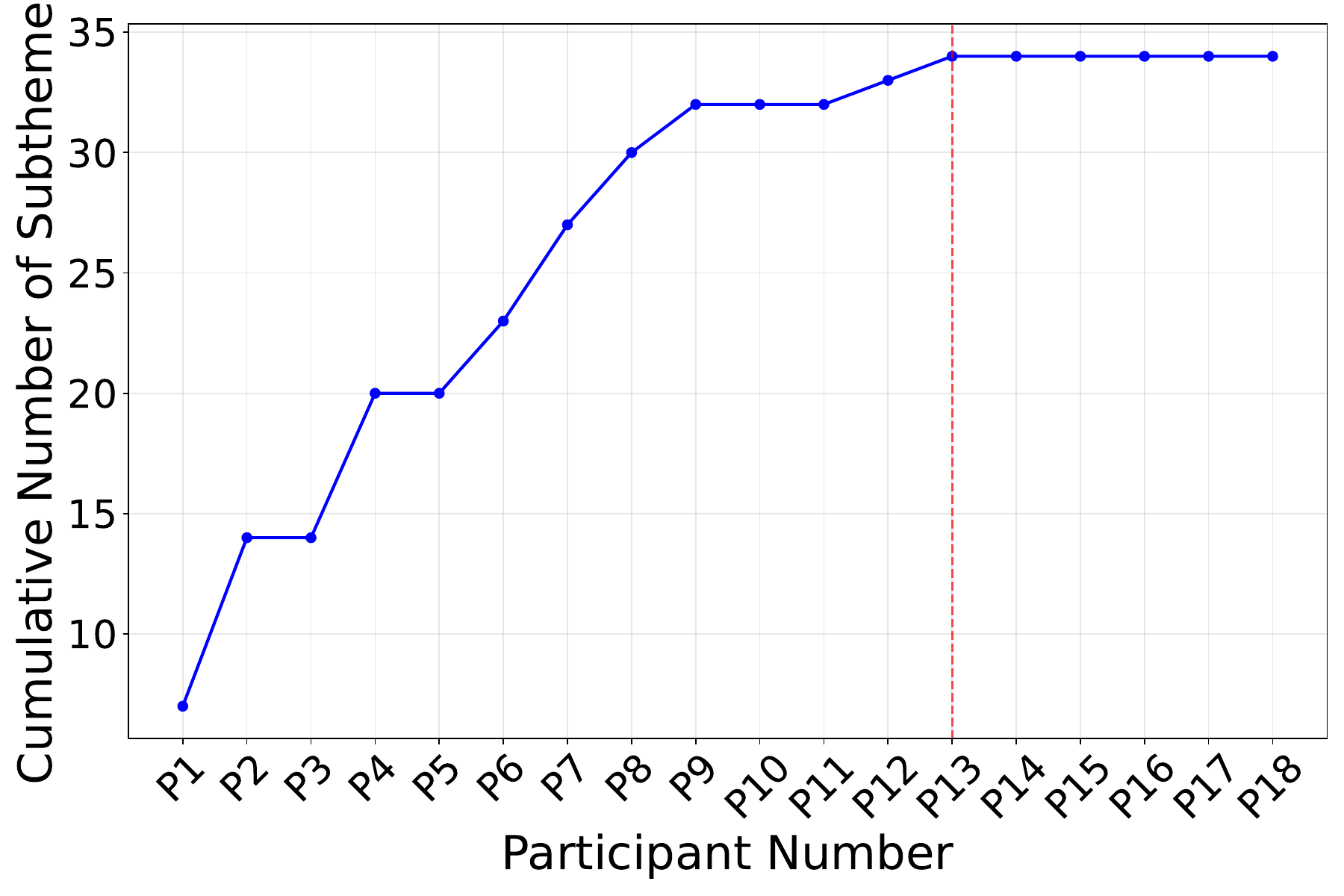}
    \caption{
    Saturation curve for interview participants. The cumulative number of subthemes saturated after P13.
    }
    \label{fig:saturation-curve}
\end{figure}

To determine whether additional participants are needed, we measured data saturation. Following~\citet{guest2006many}'s recommendation, we measured saturation by tracking the cumulative appearance of subthemes in each interview. Saturation in all RQs was achieved after interview 13. Figure~\ref{fig:saturation-curve} shows the saturation curve for the number of unique subthemes.

\subsection{Method 2: Survey}

\textbf{Instrument Design.} After gathering qualitative results for the RQs, we conducted a survey to triangulate our findings with quantitative data~\citep{easterbrook2008selecting}. We designed our survey to investigate the perceived importance of addressing specific AI failures (RQ1), frequency of adopted actions (RQ2), and the agreement on the challenges (RQ3) at a subtheme granularity. We conducted five pilot studies internally with people with experience in the development of AI-based software. Based on their feedback, we refined the clarity of the questions and adjusted the survey length. Similar to the design in~\citet{aniche2018modern}, we asked the respondents to indicate their agreement and experience with questions derived from the interview analysis subthemes. Given that developers' practices depend on the specific failure they are trying to address, for RQ2, we asked about the frequency of adopting certain actions separately for each of the three main themes from RQ1. Due to the large number of practice and AI failures combinations, we selected practices mentioned by at least half of the interview participants to minimise fatigue and ensure response quality. Detailed scales are shown in Table~\ref{tab:sample-questions}. We allowed users to respond ``\textit{I do not know}'' or ``\textit{Does not apply}'' for all questions. The survey 
included 43 questions and took approximately 15 minutes to complete.

\textbf{Recruitment.} We released our survey on the Prolific platform. To filter participants without prior experience in developing AI-based software, we applied a pre-defined filter from Prolific, requiring participants to be employed in the Information Technology (IT) sector as was done in prior work~\citep{gao2023evaluating}. 

\textbf{Data Validity Checks.} We excluded participants who reported no experience in deep learning, software engineering, or in developing AI-based software. Following~\citet{schmidt2023accountability}, we asked participants to describe one of their last three AI-based software projects and their roles for the purpose of recruiting high quality participants. 

Of the 130 participants recruited, we screened out 27 based on experience questions. Regarding the participants’ project descriptions, two authors independently reviewed the authenticity of both the projects and the reported job roles, filtering out 15 participants. This resulted in 88 valid participants after recruitment. We also included two attention check questions to ensure data quality~\citep{beach1989identifying}, which required participants to select certain choices in the survey. For example, one of the attention questions instructed the participants to select the choice of ``neutral'' as the response. Participants who failed either of the attention checks were considered invalid. This further excluded two responses, resulting in a final total of 86 valid survey submissions.

\textbf{Demographics.}
Among these subjects, 51\% had built 1-5 AI-based software projects using PTMs, 26\% built 5-10, 13\% built 10-25, and 12\% built more than 25.
For deep learning experience, 47\% had 1-2 years, 38\% had 3-5 years, 9\% had 6-10 years, and 6\% had more than 10 years.
For software engineering experience, 40\% had 1-2 years, 21\% had 3-5 years, 21\% had 6-10 years, and 20\% had more than 10 years.

\textbf{Analysis.} Similar to previous studies~\citep{gao2023evaluating}, we converted the responses to a scale of 1 to 4 for RQ2, and a scale of 1 to 5 for RQ1 and RQ3 to measure the frequency of adoption for practices, and perceived importance of each AI failure scenario and challenge, respectively. To identify whether developers adopt strategies differently when addressing different types of AI failures, we applied the Wilcoxon signed-rank test~\citep{rey2011wilcoxon} with Bonferroni correction~\citep{weisstein2004bonferroni} and Cohen's \textit{d} effect size~\citep{becker2000effect} for pairs of practices under different AI failure scenarios. This test is well-suited to our data, given the ordinal nature of the scores and the dependency between paired scores, as they were provided by the same group of participants. In cases of missing data, we included only pairs where both responses were available.

\subsection{Method 3: AI Incident Analysis}

While interviews and surveys capture downstream developers' perspectives on practical AI failures, they do not reveal whether these concerns align with documented real-world failures. To address this gap, we analysed incidents from the AIID~\citep{mcgregor2021preventing}. By triangulating these perspectives, we identify potential mismatches between developer focus and real-world failure manifestations.

\textbf{Data Source.} Managed by a board of responsible AI experts, the AI incident database compiles public submissions and provides a collective memory of failings in intelligent systems~\citep{mcgregor2021preventing}.  The database indexes a collective history of harms or near-harms caused by the deployment of artificial intelligence systems in the real world. Each entry includes a description of the incident and multiple reports linked to related news coverage. The reports are sampled to ensure comprehensive coverage of each incident and to enhance the discoverability of incidents by users. To give an example, one incident involved an automated HR system that terminated employees without human oversight, highlighting risks in AI-assisted decision-making. We downloaded all 874 incidents from the AI Incident Database on 13 Jan 2025.

\textbf{Analysis.} To map the incidents to specific failures identified in RQ1, two authors independently annotated each incident based on a specific AI failure subtheme derived from RQ1, assigning one or more failure categories. After two rounds of annotation and discussion, with each round containing 50 incidents, the authors reached a Cohen's Kappa of 0.71, indicating a good level of agreement~\citep{mchugh2012interrater}. The authors continued to annotate 100 incidents per round independently until the dataset was completed, resolving disagreements as they arose.

\subsection{Characteristics and Usage of Three Data Sources}\label{sec:data-characteristics}

In this subsection, we describe the characteristics of the three data sources and how they were used to answer our research questions.

The focus on practical AI failures resulting in immediate technical harms is shared across the three data sources. We started with a qualitative analysis of the interview transcripts to derive subthemes for each RQ. After that, we formulated the subthemes that emerged from our interview data analysis into questions (\eg How important is it to mitigate risks of the software being used irresponsibly by end-users?). Therefore, we did not need to provide a high-level theoretical definition of AI failures to our survey participants, but only specific questions that were derived from the analysis of the interviews. Similarly, we used the subthemes derived from the interviews to annotate the incidents in the AI incident database to identify which specific AI failure categories each incident exemplifies.

To assess the relevance and coverage of the three data sources, we compared the demographics of participants recruited for the interview and survey. We also compared the application domain and safety criticality of the applications that our interview participants and survey participants worked on, along with the applications mentioned in the AI incident database. Regarding the demographics of the recruited participants, interview participants had an average of 5.6 years of SE experience and 4.1 years of DL experience, with a standard deviation of 5.9 and 2.7, respectively. The median experience among interview participants was 4 years in SE and 3.5 years in DL. For the survey participants, we collected their years of experience in the form of intervals. Therefore, we take the middle point of each interval for representation to facilitate comparison. The survey participants had on average 5.6 years of SE experience and 3.6 years of DL experience (standard deviations of 3.1 and 2.5, respectively), with a median of 4 years for both SE and DL experience. In other words, the experience in SE is comparable between the two groups, while interview participants have slightly more years of experience than survey participants in DL.

Regarding the application domain and safety criticality of the developed applications, Table~\ref{tab:summarisation-data-source} summarises the application domains referenced by both interview and survey participants, along with the criticality of the systems they described. The table also includes the corresponding information from the records in the AI incident database. As some participants described software combining multiple components, such as Computer Vision (CV) and Natural Language Processing (NLP), and some survey responses or AI incident descriptions did not clearly specify which models were reused, the totals for CV and NLP applications may not sum to 100\%. Each mentioned system was annotated as either safety-critical or non-safety-critical, following the definition by~\citet{knight2002safety}, where a safety-critical system is one whose failure could result in loss of life, significant property damage, or environmental harm. During annotation, we adhered to this guideline by focusing on the system’s originally intended purpose rather than the specific context of use. For example, failures in autonomous vehicle systems are considered safety-critical because their intended purpose inherently involves risks to people. In contrast, AI game agents' inappropriate behaviours cause no direct harm to individuals, and are thus not classified as safety-critical.

Table~\ref{tab:summarisation-data-source} shows that NLP-based applications are discussed by over 70\% of the participants in both the interview and survey, while CV-based applications are discussed less. The percentage of safety-critical systems fluctuated around 10\% for the two data sources. In comparison, there were more CV-based applications mentioned in the AI incident database, and safety-critical systems were also more common there, at around 20\%.

In summary, the three data sources share a consistent interpretation of AI failures, with relatively few cases involving safety-critical systems, suggesting that the data primarily represent general-purpose AI applications rather than safety-critical domains. Although the participants’ expertise levels varied between interviews and survey, the applications they discussed were similar in terms of domain and safety criticality. In the AI incident database analysis, the application domains show more variability; however, the majority of the software applications were also not safety-critical.

\begin{table}[]
    \centering
    \caption{
    Summarisation of application and criticality of three data sources}
    \resizebox{\columnwidth}{!}{%
    \begin{tabular}{l c c |c }
    \toprule
        \textbf{Data Source} & \textbf{CV application} & \textbf{NLP application} & \textbf{Safety-Critical System} \\
        \midrule
         Interview & 37.5\% & 81.3\% & 12.5\% \\
         Survey & 15.1\% & 70.9\% & 9.3\% \\
         AI-incident database & 43.0\% & 38.1\% & 20.0\% \\
         \bottomrule
    \end{tabular}}
    
    \label{tab:summarisation-data-source}
\end{table}

In~\cref{sec:results} we provide the results to answer each of the RQs. For each RQ, we first present the qualitative analysis results from the interview participants and then quantitative findings from the survey participants. Because the AI incident database illustrates real-world failures corresponding to the identified AI failure themes, it is discussed in \cref{sec:concerns} as part of RQ1. In~\cref{sec:implications}, we structure the discussion around different stakeholder groups. The qualitative results from interviews form the foundation, while survey scores provide insight into how downstream developers perceive the importance of various concerns, the frequency with which they adopt practices, and the severity of challenges. Additionally, our analysis of the AI Incident Database serves as a proxy to assess whether these concerns are being adequately addressed in practice.

\section{Results}\label{sec:results}

\begin{figure}
    \centering
    \includegraphics[width=\linewidth]{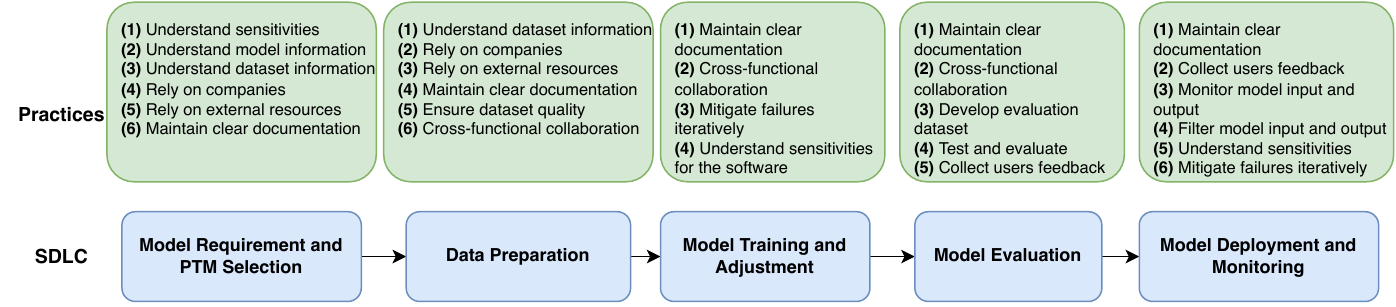}
    \caption{
    Practices mapped onto the SDLC reveal that developers frequently overlook early-stage risk assessment, concentrating instead on iterative testing during later development phases.
    }
    \label{fig:practice-process}
\end{figure}

\begin{figure}[ht!]
    \centering
    \includegraphics[width=\linewidth]{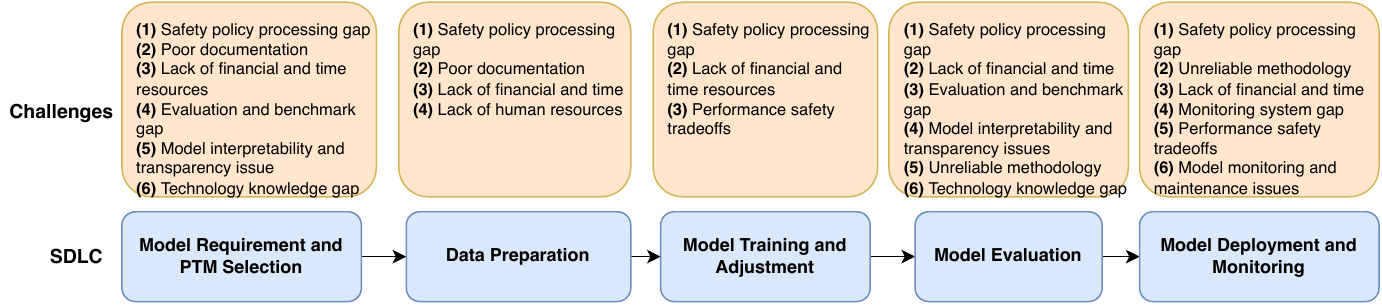}
    \caption{
    Challenges mapped onto the SDLC highlight that policy and resource constraints persistently hinder developers across all lifecycle stages, whereas technical hurdles remain phase-specific.
    }
    \label{fig:challenge-process}
\end{figure}

In this section we present our results, organised by research question.
We begin with two summary illustrations.
Table~\ref{tab:thematic-results} summarises the interview, survey and AI incident database results. Additionally, to aid interpretation and visualisation, Figure~\ref{fig:practice-process} and Figure~\ref{fig:challenge-process} offers an illustrative overview of how the subthemes of practices and challenges align with different stages of the SDLC. As we conducted interviews across three development stages, we kept notes during the coding process regarding where the discussion happened in the SDLC to construct the figure. The allocation of subthemes in the model requirement and PTM selection, as well as model deployment and monitoring were objective as our interview structure contains these stages explicitly. The intermediate stages were discerned and allocated accordingly based on the discussion context.

\begin{table}
\centering

\begin{adjustbox}{angle=90, center}
\begin{minipage}{\textheight}  
\centering

\caption{Summary of results: sub-themes and frequencies by interview participants, survey respondents, and AI incidents. The \textbf{bolded phrases} in sub-themes indicate labels used in the survey results (\cref{fig:RQ1-survey,fig:RQ2-survey,fig:RQ3-survey}). The survey score column shows average ratings (5-point scale for RQ1 and RQ3; 4-point scale for RQ2), with detailed distributions is shown in~\cref{fig:RQ1-survey,fig:RQ2-survey,fig:RQ3-survey}. }

\resizebox{\textheight}{!}{%

\begin{tabular}{cllrlr}
\toprule
                    \textbf{RQ} & \textbf{Theme}                                               & \textbf{Subtheme}                                                  & \textbf{\# Participants}  & \textbf{Survey Score (Avg. \& Med.)} & \textbf{\# AI Incidents}\\
\midrule

\multirow{8}{*}{Concerns} & \multirow{3}{*}{Data Concerns}                       & Failure risks in \textbf{fine-tuning dataset}                       &         7 (43.8\%)    & 4.56 | 5 (Range 1--5) & 66 (7.6\%) \\
                     &                                                      & Failure risks in \textbf{pre-training dataset}                       &     5  (31.3\%) & 4.52  $|$ 5& 4 (0.5\%) \\
                     &                                                      & Failure risks in \textbf{data management}                           &      4 (25.0\%)  & 4.35 $|$ 5 & 27 (3.1\%) \\
                     \cmidrule{2-6}
                     & \multirow{3}{*}{Model Behaviour Concerns}            & Model output containing harmful, biased or \textbf{inappropriate content} &       13 (81.3\%)  & 4.24 $|$ 5  & 283 (32.4\%) \\
                     
                    & & \textbf{Data leakage} and privacy breaches from the model          &      8 (50.0\%)  & 4.55 $|$ 5& 18 (2.1\%) \\
                     
                     &                                                      & \textbf{Reliability} and robustness of the model                   &         8 (50.0\%)   & 4.27 $|$ 4& 440 (50.3\%) \\
                     \cmidrule{2-6}
                     & \multirow{2}{*}{Misuse and Exploitation Concerns} & \textbf{Malicious use} of the software                             &        5 (31.3\%)    & 4.35 $|$ 5 & 242 (27.7\%) \\
                     &                                                      & \textbf{Irresponsible} and unintended \textbf{use} of the software          &  6 (37.5\%) & 4.21 $|$ 4 & 270 (30.9\%) \\  

\midrule \midrule

\multirow{15}{*}{Practices} & \multirow{3}{*}{\begin{tabular}[c]{@{}l@{}}Deeper understanding and assessment \\ for failure risks\end{tabular}}         & Understand model information                                        &      5 (31.3\%)    & ------ & ------ \\  
&&    Understand dataset content                                          &        3  (18.8\%) & ------   & ------\\

                      &                                                                                  & Understand sensitivities for the software                           &      3 (18.8\%)   & ------ & ------    \\ \cmidrule{2-6}
                      & \multirow{4}{*}{Implementing technical safeguards}                            & \textbf{Mitigate failures} by \textbf{iteratively} refining models and systems         &    11 (68.8\%)   & 3.04 $|$ 3 (Range 1--4)  & ------ \\  
                        &                                                                                  & \textbf{Ensure dataset quality}                                              &   9 (56.3\%)   & 3.25 $|$ 3 & ------  \\ 
                     &                                                                                  & Filter model input and output                                       &    7 (43.8\%)   & ------ & ------     \\  
                      & & Follow \textbf{security practices} in development of AI systems              &   9 (56.3\%)   & 3.24 $|$ 3 & ------ \\

                      \cmidrule{2-6}
                      & \multirow{2}{*}{\begin{tabular}[c]{@{}l@{}}Maintaining documentation and expert\\  consultation\end{tabular}} & \textbf{Maintain} clear \textbf{documentation}                                        &   10 (62.5\%)   & 2.93 $|$ 3   & ------ \\ 
                      &                                                                                  & Cross-functional \textbf{collaboration} for AI failure mitigation              &   9 (56.3\%)  & 3.09  $|$ 3 & ------\\ 
                      \cmidrule{2-6}
                      & \multirow{4}{*}{\begin{tabular}[c]{@{}l@{}}Regular monitoring and assessment \\ for potential failures\end{tabular}}           & \textbf{Develop evaluation dataset} for AI model failures                      &   10     (62.5\%) & 3.01 $|$ 3 & ------   \\ 
                       &                                                                                  & \textbf{Test and evaluate} AI model failures                                   &    14 (87.5\%)     & 3.07 $|$ 3  & ------ \\ 
                      & & Monitor model input and output                                       &       7 (43.8\%) & ------ & ------    \\ 
                      
                      &                                                                                  & \textbf{Collect users' feedback} regarding AI failure risks                 &   9 (56.3\%)   & 2.86 $|$ 3 & ------ \\  
                      \cmidrule{2-6}
                      & \multirow{2}{*}{Relying on external parties and resources}                       & Rely on companies to follow AI failure prevention practices                      &   8 (50.0\%)     & ------ & ------     \\  
                      &                                                                                  & Rely on external resources for model quality and content integrity &   4 (25.0\%)   & ------ & ------     \\ 
                      \midrule \midrule
\multirow{11}{*}{Challenges}                       & \multirow{2}{*}{Infrastructure gaps}                                                                                                             & Evaluation and \textbf{benchmark gap}                                                             &           6 (37.5\%)             &    4.09 $|$ 4 (Range 1--5)    & ------  \\
                                            &                                                                                                                                                  & \textbf{Monitoring }system \textbf{gap}                                                                    &                2   (12.5\%)        & 4.20 $|$ 4 & ------       \\\cmidrule{2-6}
                                            & \multirow{3}{*}{Technical and model-related difficulties}                                                                                        & Model interpretability and \textbf{transparency issue}                                            &            4  (25.0\%)  & 4.17 $|$ 4   & ------          \\
                                            &                                                                                                                                                  & Performance \textbf{failure mitigation tradeoffs}                                                             &           3 (18.8\%)   &    3.72 $|$ 4     &  ------        \\
                                            &                                                                                                                                                  & Model monitoring and \textbf{maintenance issue}                                                  &            4 (25.0\%) & 4.08 $|$ 4   & ------                \\ \cmidrule{2-6}
                                            & \multirow{2}{*}{Lack of mature process and methodology}                                                                                          & Risk management \textbf{policy} interpretation \textbf{gap}                                                             &            6 (37.5\%)  & 4.01  $|$ 4   & ------              \\
                                            &                                                                                                                                                  & \textbf{Unreliable methodology}                                                                   &                          7 (43.8\%)   & 4.13 $|$ 4   & ------  \\ \cmidrule{2-6}
                                            & \multirow{2}{*}{Resource constraints}                                                                                                            & Lack of \textbf{financial} and time resource                                                      &                   6 (37.5\%)      & 4.00 $|$ 4  & ------     \\
                                            &                                                                                                                                                  & Lack of \textbf{human} resources                                                                  &                    5 (31.3\%)              & 4.26 $|$ 4.5 & ------ \\ \cmidrule{2-6}
                                            & \multirow{2}{*}{\begin{tabular}[c]{@{}l@{}}Barriers to technical understanding \\ and information access\end{tabular}}                           & Technology \textbf{knowledge gap}                                                                &                   3 (18.8\%)    & 4.21 $|$ 4  & ------        \\
                                            &                                                                                                                                                  & \textbf{Poor documentation}                                                                       &  4 (25.0\%) & 4.26 $|$ 4 & ------\\
\bottomrule
\end{tabular}}

\label{tab:thematic-results}
\end{minipage}
\end{adjustbox}

\end{table}

\subsection{RQ1: Practical AI Failure Concerns}\label{sec:concerns}

\begin{figure} 
    \centering
    \includegraphics[width=0.9\columnwidth]{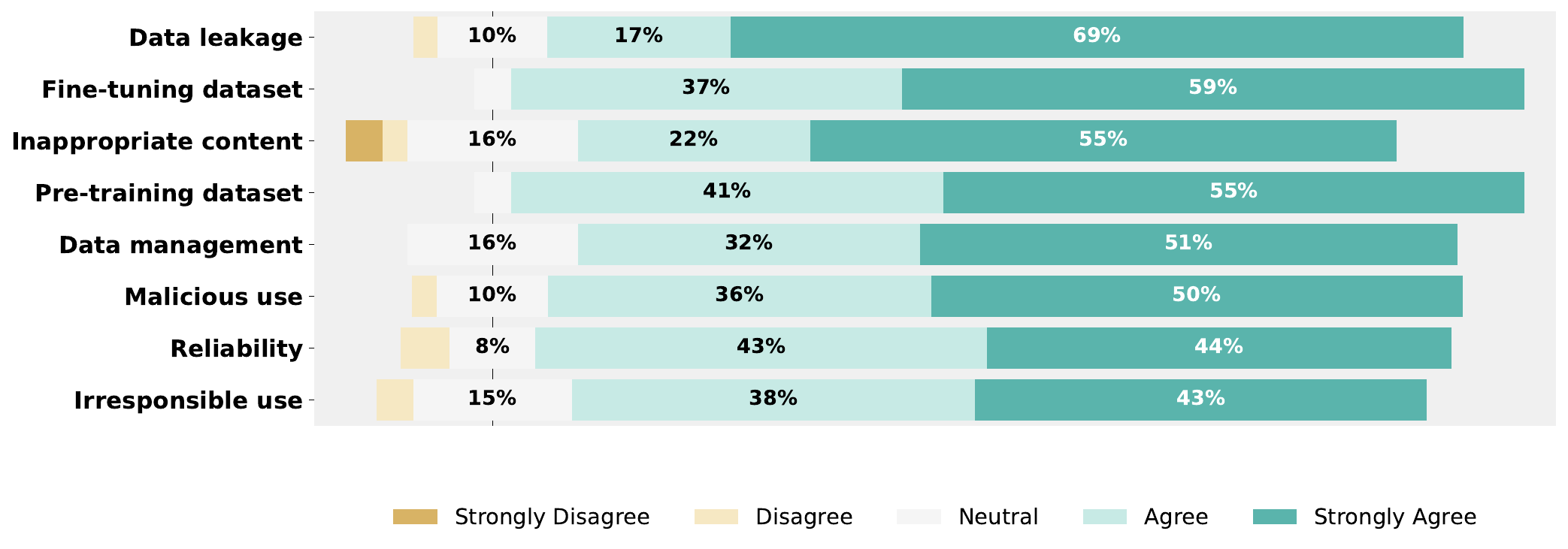}
    \caption{
    Survey results on importance of AI failure concerns.
    The full details about the distribution are in our replication package in~\cref{sec:replication}. Percentages may not sum to exactly 100\% due to rounding. Data labels are omitted where the corresponding segment is too small to fit the text.}
    \label{fig:RQ1-survey}
\end{figure}

Developers' AI failure concerns are three themes, namely (1) data concerns, (2) model behaviour concerns, and (3) misuse and exploitation concerns. Figure~\ref{fig:RQ1-survey} illustrates the RQ1 survey results. 

\textbf{Theme 1: Data Concerns} relates to data content and management in AI software development. The most discussed subtheme, \ul{\textit{\mbox{failure} risks in fine-tuning datasets}}, raised by seven participants, highlights concerns about biased, inappropriate, or personally sensitive content in fine-tuning datasets, which could impact model behaviour~\citep{gong2023survey}. P11 reflected the privacy aspect by stating: \textit{``If I'm training any sort of image generation models, ..., I can't just use someone's photo...''}. In our survey, 97\% respondents rated fine-tuning dataset content as   ``very important'' or ``extremely important''. 

\ul{\textit{\mbox{Failure} risks in pre-training datasets}} impacts the PTM behaviour. Similar to fine-tuning datasets concerns, practitioners perceive inappropriate, biased, and privacy-sensitive content within pre-training datasets as catalysts for AI failures. However, while fine-tuning datasets can be later modified, pre-training dataset content serves as a proxy for understanding the unintended risks of the PTM. Some participants also noted detecting backdoor data. In our survey, 96\% rated checking pre-training dataset content as ``very important'' or ``extremely important''.

\ul{\textit{\mbox{Failure} risks in data management}}, concerns risks in storing, accessing, and transferring AI artefacts. For example, P7 noted ``\textit{whether the storage method is safe, and also this includes storage for the data and the model}''. In our survey, 84\% respondents consider it ``very important'' or ``extremely important'' to establish safe data management process. 

\textbf{Theme 2: Model Behaviour Risks} refers to direct technical risks related to model output. \ul{\textit{Model output containing harmful, biased, or inappropriate content}} is the most mentioned subtheme, highlighted by 13 participants. Biased content relates to gender, race, and other factors; harmful content includes offensive or violent outputs; and inappropriate content involves sensitive or erotic topics~\citep{gao2024documenting}. For example, P15 noted that \textit{``some of the retrieved images could be inappropriate to many users.''} In our survey, 78\% considered it as ``very important'' or ``extremely important''. 

\ul{\textit{Data leakage and privacy breaches from the model}} occur when outputs resemble private data from fine-tuning or when third-party APIs collect sensitive information. For example, P7, who worked on an AI-based application in the banking domain, noted that customers particularly prioritised privacy when using such software. 

\ul{\textit{Reliability and robustness of the model}} concern performance for AI systems in critical domains and its robustness to varied inputs. In the context of LLMs, participants also associate this with hallucinations~\citep{zhang2023siren}. This concern is raised particularly when the downstream application is in critical domains, such as healthcare, or the AI-component works as an essential role in the software pipelines. Data leakage and privacy breaches, along with model reliability and robustness, received strong support in the survey, with 87\% respondents rating them as ``very important'' or ``extremely important''.

\textbf{Theme 3: Misuse and Exploitation concerns} relates to how downstream users engage with AI-based software. Six participants highlighted concerns about \ul{\textit{irresponsible and unintended use of the software}} including users probing models with unsafe questions, using the software beyond its intended scope, or over-relying on model outputs. For example, P18 reflected on cases when AI-based software used for medical purposes despite not obtaining certification for such use purposes and not under regulation.   Meanwhile, \ul{\textit{malicious use of the software}} associates with downstream users deliberatively leveraging the AI-based software to steal training data and model information, or generate harmful content using techniques such as jailbreaking~\citep{wei2024jailbroken}. In our survey, 81\% and 87\% respondents rated irresponsible and unintended use, and malicious use of software,  as ``very important" or ``extremely important", respectively. However, irresponsible and unintended use of software received the lowest survey score among all concerns at 4.21.

\textbf{AI incident database analysis result:} Our AI incident database analysis highlights real-world AI failures from a user-facing perspective. This helps assess whether downstream developers' awareness and practices are sufficient. Table~\ref{tab:thematic-results} displays the results.
Model reliability and robustness failures are most frequent (50.3\% of cases). A representative case involves autonomous driving algorithms running red lights or misidentifying objects, leading to severe consequences.  32.4\% of the incidents involve inappropriate, harmful, or biased outputs. Misuse and exploitation failures are also frequent. Irresponsible or unintended use cases or over-reliance on the software accounts for 30.9\% of the incidents, making it the third most frequent type of incident. A typical case for this type of failure involves over-reliance on the provided output even without human intervention, such as automating employee layoffs without human oversight. Malicious use makes up 27.7\%. Since the AI incident database is user-facing, data-related failures are less reported: incidents related to fine-tuning dataset content, pre-training dataset content, data management, and data leakage and privacy breaches, account for 7.6\%, 0.5\%, 3.1\% and 2\% of cases, respectively.

To interpret these data, we compare them to the interview and survey results.
``\textit{Model reliability and robustness}'' and ``\textit{Output containing harmful, biased or inappropriate content}'' are the most commonly discussed subthemes in interviews. Their reported failure cases ranked second and first, respectively, suggesting that despite high awareness of these potential failures, issues in mitigation approaches may exist. In contrast, subthemes under \textit{misuse and exploitation concerns} were less mentioned in interviews, with \textit{irresponsible and unintended use of the software} receiving the lowest importance score. However, these failure types still account for a significant portion of incidents, indicating a lack of attention from downstream developers.

\begin{tcolorbox}[left=1pt, top=1pt, right=1pt, bottom=1pt] \textbf{Highlights}: Developers' AI failure concerns focus on data, model behaviour, and misuse/exploitation concerns.  Meanwhile, the AI incident database reveals model reliability and robustness issues as the most frequent failures, accounting for over half of cases.
\end{tcolorbox}

\subsection{RQ2: Practices When Approaching AI Failures}

\begin{figure}
    \centering
    \includegraphics[width=0.9\linewidth]{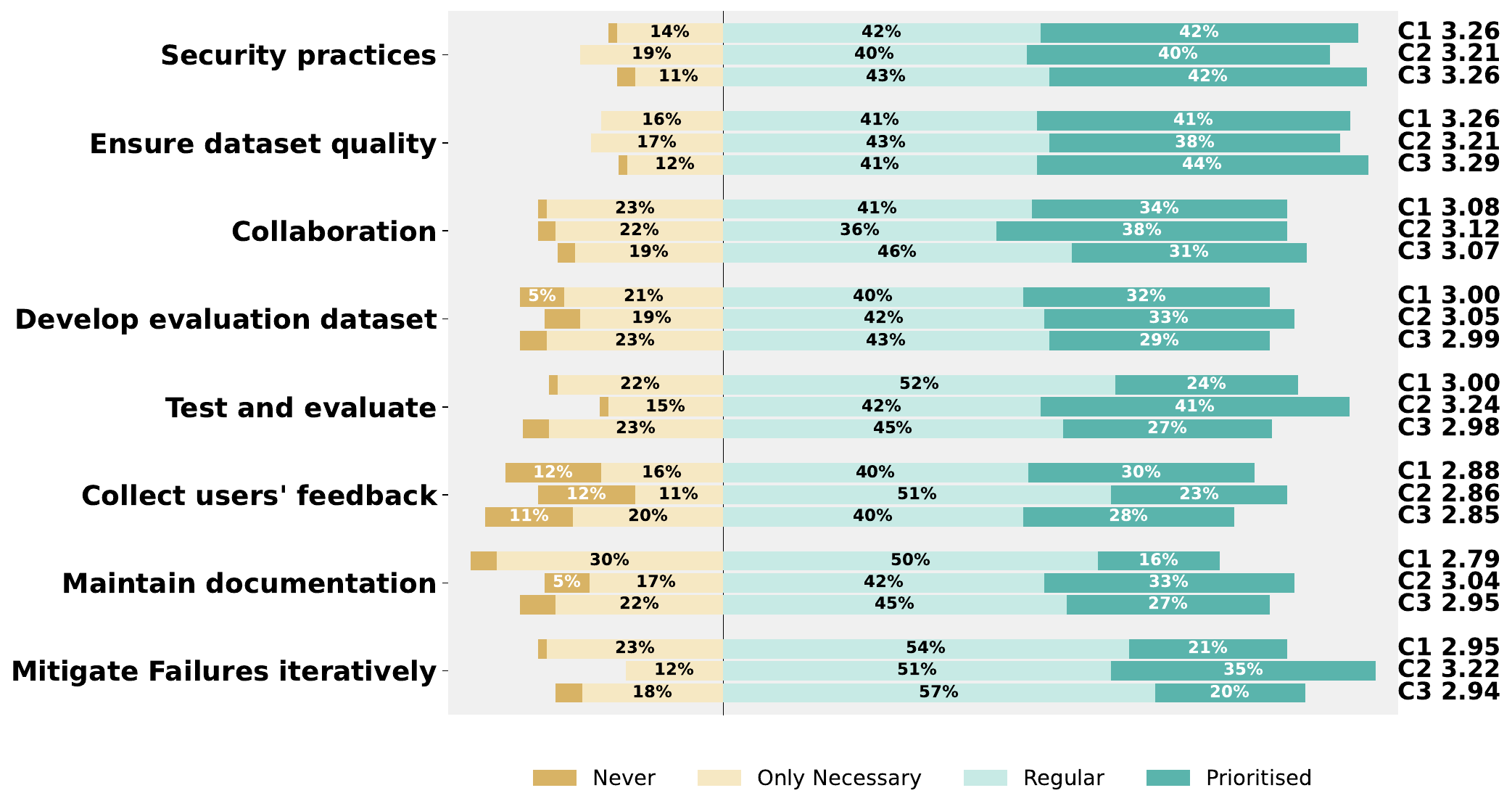}
    \caption{Survey results on frequency of failure mitigation practices. Each practice includes three rows representing the reported frequencies when addressing the three failure concerns identified in RQ1 (C1: data concerns, C2: model behaviour concerns, and C3: misuse and exploitation concerns). For example, the first group of bars indicates how frequently participants adopt "Security practices" when mitigating C1, C2, and C3, respectively, along with the average frequency score for each. Full details about the distribution are in our replication package~\cref{sec:replication}. Percentages may not sum to exactly 100\% due to rounding. Data labels are omitted where the corresponding segment is too small to fit the text.}
    \label{fig:RQ2-survey}
\end{figure}

The practices we observed span  five themes:
  (1) deeper understanding and assessment for failure risks;
  (2) regular monitoring and assessment for potential failures;
  (3) implementing technical safeguards;
  (4) relying on external parties and resources; and
  (5) maintaining documentation and expert consultation. 
  Figure~\ref{fig:RQ2-survey} quantifies the frequency of practitioners' strategy choices when approaching three practical AI failure concerns we identified in RQ1 (C1: data concerns, C2: model behaviour concerns, C3: misuse and exploitation concerns).

\textbf{Theme 1: Deeper Understanding and Assessment of Failure Risks} focuses on the preparation phase of developing AI-based software, where developers familiarise themselves with potential risks in AI-related artefacts, development processes, and use cases.  \ul{\textit{Understanding Model Information}} involves reviewing PTM information related to ethical and robustness aspects from online sources, such as research papers or benchmarks. Additionally, P8 suggested manually inspecting PTM parameter storage and loading scripts to detect malicious code, and also endorsed using the SafeTensors format~\citep{huggingfaceSafetensors} PTMs over other binary formats.

For \ul{\textit{understanding dataset content}}, developers examine both the pre-training dataset content regarding sensitive and inappropriate content, as a proxy for the PTM behaviour, and fine-tuning dataset to assess data collection quality and potential risks in subsequent AI-based software processes. These manual checks involve reviewing subsets of dataset content or descriptive dataset information from sources such as Hugging Face data cards. For example, P10 mentioned ``\textit{I will check GitHub or Hugging Face page, and randomly select some content to manually go through the data}''. 

Additionally, \ul{\textit{understanding sensitivities for the software}} involves identifying the target users of the AI-based software and specific sensitive information for them. For example, P6 reflected on the importance of identifying target users: ``\textit{We need to see who the marginalised groups are—people who could be subject to some form of discrimination}''.

\textbf{Theme 2: Implementing technical safeguards} involves adopting methods to mitigate potential AI failures during development. \ul{\textit{\mbox{Mitigate failures} by iteratively refining models and systems}} requires developers to adjust models, evaluate potential failure factors, and iterate until satisfactory results are achieved. The fine-tuning process may be examined, as P13 mentioned: ``\textit{first we tried to see if we did something wrong in the fine-tuning process}''. This also includes examining fine-tuning processes, switching models or checkpoints, and retraining on improved datasets. For example, P17 fine-tuned with a more diverse accent dataset to enhance speech recognition performance for rural accents, reducing the risk of unfair or harmful outcomes caused by systematic recognition errors in underrepresented populations. For LLMs, practitioners also refine prompts to prevent undesired content generation. P13 reflected: ``\textit{in terms of politeness that I mentioned, sometimes we try to change the prompt we are using for the model to prevent it from talking things we don't want to do}''. Our survey revealed significant differences in responses depending on the AI failure context. Developers performed iterative model and system improvement significantly more frequently when addressing model behaviour risks (3.24) compared to misuse and exploitation (2.98) concerns, with a p-value of 0.004 and an effect size of 0.33. These results suggest that different AI failure concerns have a medium effect on developers' frequency of adopting iterative model and system improvement. Only 13\% of the participants put low priority on this practice when dealing with model behaviour concerns, which is lower compared to the scenarios when dealing with data or misuse and exploitation concerns.

The previous efforts rely on the actions to \ul{\textit{ensure dataset quality}}. One approach is to incorporate higher-quality data into fine-tuning dataset for the prioritised AI failure concerns. For example, since P17's speech recognition product targets children in multiple rural areas, they partnered with local organisations to gather diverse accents, enriching their fine-tuning data for model improvement. Filtration methods are also used to remove biased, violent, or personally sensitive content in fine-tuning datasets. For example, one participant (P7) described masking customers’ sensitive information before model training to prevent privacy breaches in AI-based applications. These filtration methods involve manual annotation by human reviewers or automated filtering. Also, sampling and data engineering techniques are employed to enhance datasets to provide better inputs aligned with the prioritised AI failure concerns. Our survey shows this practice had the highest average score of 3.25, aligning with the high importance placed on data concerns. More than 80\% of the survey participants consistently adopt this strategy as a regular or priority practice when handling all the mentioned AI failure concerns.

\ul{\textit{Filtering model input and output}} acts as an external safeguard against AI failures, using both manual rules and automated filters. For example, P15 applied filters to detect inappropriate content, while P12 used regular expressions to remove privacy-sensitive information from model-generated text. For LLMs, P11 highlighted using Llama Guard~\citep{inan2023llama} as a post-processing step for model output. 

\ul{\textit{Following security practices in the development of AI systems}} relates to particular processes needed in the development. Several participants highlighted local solutions, such as on-premise data storage and training pipelines to prevent data leakage. For example, P12 addressed data storage concerns by stating: ``\textit{The data leakage can happen and to resolve it, ... on premise the installation of databases}''. Other security measures include encrypting data and models, maintaining backups, and selectively publishing checkpoint versions. Based on our survey data, this subtheme is adopted frequently, receiving an average score of 3.24, and it was considered a regular to priority practice by more than 80\% of the participants across all AI failure concern contexts.

\textbf{Theme 3: Maintaining documentation and expert consultation} contains several procedural efforts. \ul{\textit{Maintain clear documentation }} refers to the process of providing documentation for both internal review and downstream users to ensure transparency. P9 mentioned that all processes, including decisions on PTM selection, model robustness, and efforts to mitigate risks, need to be clearly documented for company's audit. Participants also made efforts in recording model evaluations for various potential AI failures, providing guidelines for responsible interaction with the software, and acknowledging potential technical limitations in the documentation. Additionally, documentation on disclaimers was mentioned, with P12 highlighting its use as a way to bypass responsibility for potential AI failures. Our survey respondents reported varying and slightly lower frequencies of documentation practices across different concerns: data (2.79), model behaviour (3.04), and misuse and exploitation (2.95). Overall, documentation ranked among the least frequently adopted mitigation practices, with around 24\% to 34\% of the participants de-prioritising this task. 

Meanwhile, \ul{\textit{cross-functional collaboration for \mbox{AI failure mitigation}}} involves the cooperative efforts to address potential AI failure concerns, including discussions on risks, collaboration among multiple teams or organisations to collect datasets, test AI models for potential failures, deploy software securely, and maintenance over time. Moreover, given the expertise gap, AI software developers collaborate with domain experts throughout the development process to analyse risks, build evaluation ground truth datasets, and conduct further evaluations. For example, when handling sensitive customer data, P13 reflected: ``\textit{I'm not an expert in that field [Data Privacy Compliance]. I just talk with the legal team or the compliance team and if they say no, I just stop ... we are working constantly with that team}''. This subtheme received an average score of 3.09, with fewer than 30\% of participants indicating it was a low priority, suggesting broad adoption among practitioners.

\textbf{Theme 4: Regular monitoring and assessment for potential failures} involves evaluating failure risks before release and continuous monitoring systems after deployment. A key step is to \ul{\textit{develop an evaluation dataset for AI model \mbox{failures}}}, with diverse efforts observed, from simply using open-source data to crowdsourcing or even expert consultations. One way to construct evaluation dataset is to reuse existing evaluation datasets on open-source platforms, as P4 mentioned collecting testing documents related to sensitive topics on GitHub. Efforts are also spent refining existing datasets by reformulating certain details. For instance, P6 described adapting culturally American-centric question-answer pairs to reflect the traditions of other cultures when evaluating models. When assessing model's reliability, robustness, or data leakage, practitioners often use ground-truth or production data. For hallucination evaluation, P16 reflected: ``\textit{the customer support case, ..., we will have transcripts of customer support conversations}''. Last, P13 also exploit the possibility of using LLMs to generate test cases for evaluation: ``\textit{we have used some generative AI models like GPT to help us generate some questions}''.
Our survey results showed this practice received an average score of 3.01 for its frequency of adoption, with 73\% to 76\% of participants reporting it as a regular or prioritised activity, indicating it is widely used for AI failure assessments.

After that, developers would \ul{\textit{test and evaluate AI model \mbox{failures}}}. Diverse methods were observed among the participants. Some efforts are preliminary, involving only probing the model with small samples and manual inspection, such as  P4:``\textit{we randomly select some questions and test the LLM to see if it response with safe answers}''. Automatic evaluation using traditional metrics such as Rouge~\citep{lin2004rouge} was also mentioned. In particular, LLM-as-judge~\citep{zheng2023judging} is used by developers to evaluate specific AI failures such as hallucinations, bias, and harmful generation, as reflected by P16. Also, a hybrid approach combines manual and automated effort, as P18 described for privacy evaluation: ``\textit{then flagging low scoring examples in a development set for ... human evaluation}''. Our survey indicates that developers conduct AI failure testing and evaluation significantly more frequently when addressing model behaviour risks (3.22) compared to data concerns (2.95) and misuse and exploitation concerns (2.94), with p-values of 0.007 and 0.004 and effect sizes of 0.40 and 0.41, respectively. 16\% of the participants put low priority on this practice when dealing with model behaviour concerns, while this number increased to 23\% and 27\% within the context of data concerns and misuse and exploitation concerns, respectively. These results suggest that different AI failure concerns have a moderate effect on the frequency with which developers test and evaluate for AI failures.

After deployment, developers made continuous efforts to gauge AI safety concerns in the AI-based software. Regarding \ul{\textit{monitor model input and output}}, developers actively track content from both model inputs and outputs to detect any violations of AI safety. For example, P11 noted: ``\textit{get notification of every prompts that's been done to my model and I see the output as well}''. Moreover, continuous evaluation of the model’s reliability is conducted by periodically measuring key metrics. This allows practitioners to identify model behaviour drift and make necessary adjustments accordingly. 

Regarding \ul{\textit{collect users' feedback regarding \mbox{AI failure risks}}}, practitioners monitor product reviews across various channels, such as app store reviews, and sometimes conduct pre-release user tests. However, this subtheme received the lowest average score of 2.86, with 12.8\% of respondents reporting that they never engaged in this practice. This suggests a relative lack of effort among AI developers in collecting user feedback on AI failures, partly because it is often perceived outside the typical scope of software engineering responsibilities. However, even among open-source developers who oversee entire projects, several mentioned releasing their software “as is” and only rarely revisiting user feedback

\textbf{Theme 5: Relying on external parties and resources} describes developers' dependence on external components like models, datasets, and libraries instead of addressing concerns themselves. In the subtheme of \ul{\textit{relying on companies to follow AI \mbox{failure prevention} practices}}, developers often trust the reputation of the companies from which they obtain AI models, datasets, and related services. For example, P1 noted ``\textit{A lot of people seem to go with Meta. You don't really need to think too much about the ethical side}''. This reliance also extends to company-provided workflows and development processes, such as automated data collection pipelines.

Meanwhile, \ul{\textit{rely on external resources for model quality and content integrity}} reflects  the dependence on popular models, libraries, and datasets. For example, P11 mentioned his trust in HuggingFace safetensor format by reflecting: ``\textit{you're bound to have less adversarial attacks compared to your open source ONNX or other formats}''.

\begin{tcolorbox}[left=1pt, top=1pt, right=1pt, bottom=1pt] 
\textbf{Highlight}: Practitioners' practices towards AI failure concerns during the development stages include (1) deeper understanding and assessment of failure risks, (2) implementing technical safeguards, (3) maintaining documentation and expert consultation, (4) regular monitoring and assessment for potential failures, and (5) relying on external parties and resources. In particular, documentation and feedback collection are less frequently adopted, and few interview participants consider understanding failure risks as an initial step.
\end{tcolorbox}

\subsection{RQ3: Perceived Challenges}
\label{sec:RQ3_challenges}
\begin{figure}
    \centering
    \includegraphics[width=0.9\linewidth]{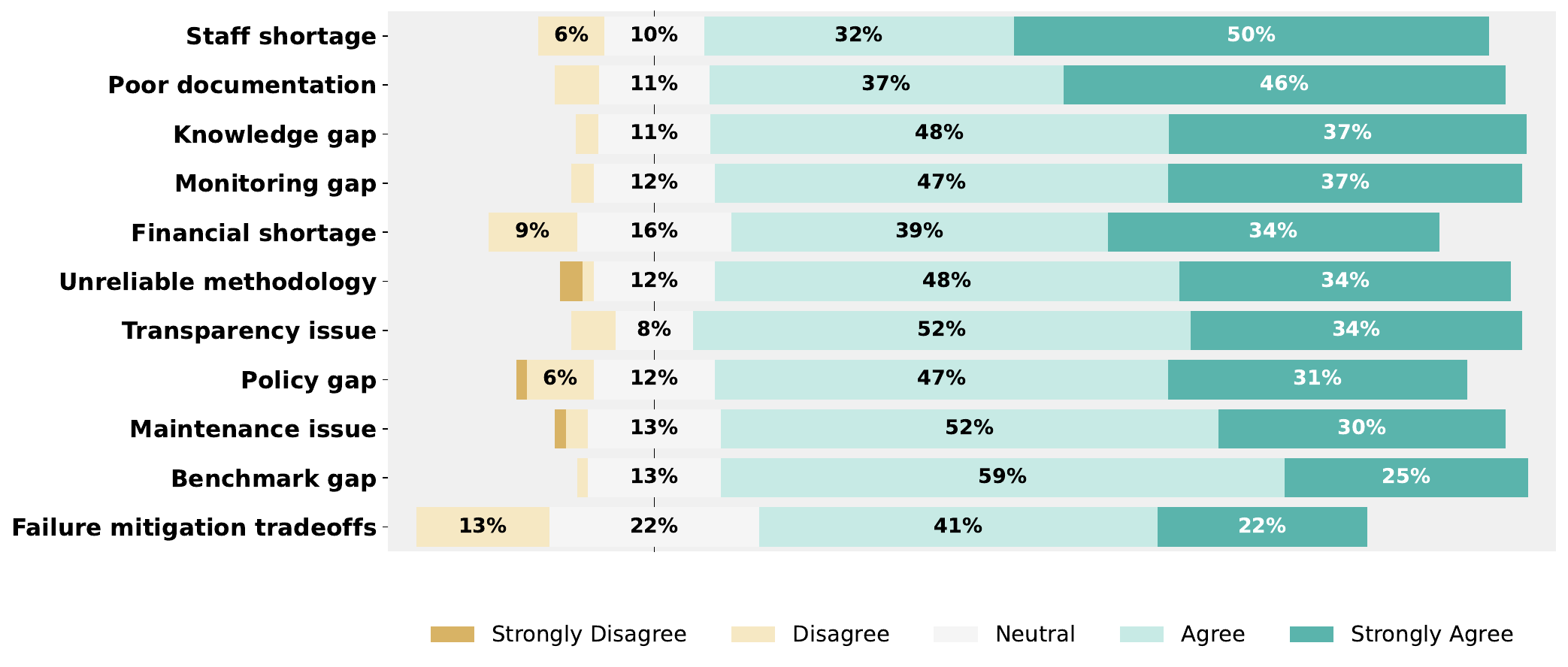}
    \caption{Survey results on agreement on AI failure challenges. For example, 82\% of the participants agreed or strongly agreed (32\% and 50\%, respectively) that staff shortage poses a challenge. The full details about the distribution is in our replication package in~\cref{sec:replication}. Percentages may not sum to exactly 100\% due to rounding. Data labels are omitted where the corresponding segment is too small to fit the text. }
    \label{fig:RQ3-survey}
\end{figure}

Five main themes of challenges when handling AI failure concerns are faced by the participants, namely (1) infrastructure gaps, (2) technical and model-related difficulties, (3) lack of mature process, (4) resources constraints, and (5) barriers to technical understanding and information access. Figure~\ref{fig:RQ3-survey} displays the survey results. 

\textbf{Theme 1: Infrastructure gaps} refers to the absence of artefacts and tools for managing AI failure. \ul{\textit{Evaluation and benchmark gaps}} highlights the lack of metrics and benchmarks for evaluation. P4 noted:  ``\textit{There is no metric to decide if the model is safe enough}". The lack of benchmark includes both an absence of standardised datasets for specific AI failure concerns and limited datasets marginalised groups such as less common cultures. In particular, P9 mentioned: ``\textit{It would be great if we have the evaluation dataset... we have to do it by ourselves, and it is very hard to control}''.

Similarly, \ul{\textit{monitoring system gaps}} reflect the absence of systems to track deployed AI software. For instance, P13 emphasised the lack of a dashboard-like monitoring system to proactively detect failure occurrences over relying on user feedback. Our survey confirms these challenges, receiving 85\%  agreement on the challenges of evaluation and benchmark gaps, as well as monitoring system gaps.

\textbf{Theme 2: Technical and model-related difficulties} includes both inherent and external challenges in addressing AI failures within model and software systems.  \ul{\textit{Model interpretability and transparency issue}} hinders model understanding, making failure-mitigation fine-tuning and testing difficult. As P12 noted: ``\textit{AI models are often black boxes, making explainability difficult, which complicates troubleshooting unexpected behaviour}''. This was the most acknowledged challenge, with 87\% agreement in the survey. 

\ul{\textit{Performance \mbox{failure mitigation} tradeoffs}} arises from accuracy loss in mitigation fine-tuning and efficiency reductions from behavioural guardrails, complicating the balance between performance and reliability. For example, P2 noted: ``\textit{There is usually a trade-off between utility, efficiency, and safety. If I want to increase safety, I often have to sacrifice other factors}''. This challenge was recognised by 64\% of survey participants, though 14\% disagreed.

The key challenge in \ul{\textit{model monitoring and maintenance issues }} is tracking behavioural drift over time and updating systems for new failure modes. As P16 noted: ``\textit{So the questions that are asked, the issues that are faced these change over time. And so what's difficult, especially for a customer facing issue is the ability to continue to update the information}''. This challenge is worsened by insufficient monitoring metrics and the lack of concrete thresholds for necessary updates. Additionally, adapting updates is difficult due to the challenge of reproducing the observed failures. 83\% of our survey participants agreed with this challenge.

\textbf{Theme 3: Lack of mature process} refers to insufficient and unreliable methods for addressing AI failures during development. \ul{\textit{\mbox{Risk management}
interpretation gap}} highlights the lack of well-defined policies and systematic approaches. For instance, P9 emphasised ambiguity in policy interpretation: ``\textit{For example, you cannot use `untrusted third-party library'. The definition is not really clear ... so sometimes we need to decide by ourselves}''. P12 noted the complexity in ensuring compliance with legal and ethical standards in multi-regulatory environments. This gap extends to the absence of standardised approaches for managing various AI failures, with P18 adding: ``\textit{there's a lack of clarity or even in my opinion, community-agreed best practices on how to do monitoring}''. In our survey, 79\% respondents agreed that the policy interpretation gaps pose a challenge. 

\ul{\textit{Unreliable methodology}} refers to the flaws in methodologies themselves, encompassing the limitations of data annotation and crowdsourcing processes~\citep{hsueh2009data}. For example, P12 noted: ``It may be correct from one person’s perspective, but someone else might consider it wrong from their own ... It (the resolving process) is very hectic, and can take months to resolve it''. Further concerns include the questionable reliability of user feedback and the inherent difficulty of comprehensively anticipating AI failure risks during development. For example, P11 noted: ``\textit{What I ended up doing is basically the problems that I see, but not something that I was not expecting them to happen}''. This challenge was also strongly supported, with 84\% of survey participants acknowledging it as a difficulty.

\textbf{Theme 4: Resource constraints} refers to the lack of essential resources. The main subtheme is  \ul{\textit{lack of financial and time resources}}, as mitigating AI failures requires extra steps, specialised expertise, and more time, creating a significant financial burden. P14 noted: ``\textit{since we are a small company, we don't have a big budget}''. This challenge was supported by 74\% of the survey respondents. 

With these additional steps to handle AI failures also comes \ul{\textit{lack of human resources}}. The process of high quality dataset curation requires human resources. Also, most developers of AI-based software are not specialised in handling varieties of AI failures, necessitating the involvement of more experts, as highlighted by P15: ``\textit{But you don't have any specialist for the AI safety}''. This challenge was agreed by 83\% of participants and received the highest score of 4.26, identified as one of the biggest challenges developers face. 

\textbf{Theme 5: Barriers to technical understanding and information access} encompasses challenges in accessing and understanding failure mitigation implementation. The subtheme \ul{\textit{technology knowledge gaps}} highlights difficulties arising from unfamiliarity with failure mitigation terminology and techniques. P2 noted the challenge of technical literacy when discussing AI model privacy metrics:  ``\textit{For our general users, I think most people they do not know the concept of privacy budgets~\footnote{Privacy budget is a concept in differential privacy that limits the amount of information an individual's data can reveal in a dataset~\citep{li2010optimizing}.}}''. This issue also affects cross-team collaboration due to inconsistencies in terminology among stakeholders~\citep{nahar2022collaboration}. This challenge is agreed by 86\% of the survey participants. 

Meanwhile, \ul{\textit{poor documentation}}  refers to incomplete or absent potential failure risks evaluation records for PTMs and datasets. P18 highlighted a critical mismatch between available and required information: ``\textit{There's often a misalignment between what benchmarks are publicly available for the particular task and domains that are actually relevant to the project}''.  This comment illustrates how existing benchmarks frequently fail to cover domain-specific needs, leaving developers without meaningful evaluation resources. This challenge received the highest average score (4.26), indicating a significant lack of documentation efforts.

\begin{tcolorbox}[left=1pt, top=1pt, right=1pt, bottom=1pt] \textbf{Highlights}:Developers face multifaceted challenges in addressing AI failures, spanning missing infrastructure, technical and methodological limitations, immature processes, resource constraints, and limited expertise and documentation. In particular, the lack of human resources and poor documentation are perceived as the biggest challenges in mitigating AI failures.
\end{tcolorbox}

\section{Discussion and Implications}\label{sec:implications}

We discuss implications for different stakeholders in this section.

\paragraph{Developers of AI-based software.} Implications are discussed following the ML development process described by~\citet{amershi2019software} and~
\citet{jiang2023empirical}, which we divide into three stages mapped onto the SDLC.

    (1) \ul{\textit{Preparation and PTM Selection stage:}}  Based on practices in Table~\ref{tab:thematic-results}, interview participants rarely reflected on deeper understanding and assessment for failure risks.
    These actions mainly occur in initial stages of development processes as shown in Figure~\ref{fig:practice-process}, suggesting insufficient attention from developers in analysing and anticipating potential failure risks from the outset. Meanwhile, PTM attributes and potential risks have been highlighted as essential considerations at the initial stage~\citep{jiang2023empirical}. Therefore, developers of AI-based software should dedicate more efforts to assessing risks and intended usage early on, as this proactive approach can reduce downstream challenges and help prevent common misuse and exploitation failures observed in the AI incident database, ultimately contributing to the goal of AI safety where the immediate harms from failure of AI components are mitigated.  One possible way to foster the early-stage reflection is through fictional design approaches. For example, fictional research abstract is found to help in anticipatory reflection on AI ethics~\citep{jaaskelainen2025anticipatory}, and such methods could potentially be adapted to encourage safety-focused and failure-prevention reflection during the PTM selection stage.
    
    (2) \ul{\textit{Model Development and Testing:}} Developers frequently adopted practices related to model development and testing, especially for the subthemes of \textit{test and evaluate AI model \mbox{failures}} and \textit{\mbox{mitigate failures} by iteratively refining models and systems} as indicated in Figure
    ~\ref{fig:practice-process}, which are found even more frequently when dealing with model behaviour concerns in AI failures (see~\cref{fig:RQ2-survey}). However, frequent failures in the AI incident database raise questions about their effectiveness. In fact, we observed varying levels of systematic and comprehensive approaches, with some participants investigating potential AI failures by randomly probing a few examples. Others invested more effort, evaluating real-time data using a combination of automated metrics and human assessment. Recent machine learning research has introduced methods to assess bias~\citep{salewski2023incontext}, toxicity contents~\citep{shaikh2022second}, reliability~\citep{wang2023on}, with various treatments~\citep{jung2024a, chen2023holistic}. However, knowledge gaps, insufficient documentation, and lack of policies hinder developers from effectively leveraging these methods. To mitigate this knowledge transfer gap, downstream developers should actively seek out new techniques. Given the rapid pace of advancements, aggregator platforms like Reddit or Twitter~\citep{aniche2018modern} can serve as valuable resources for staying updated on emerging failure mitigation solutions and further discourse on ensuring the general AI safety. However, developers should also be mindful as not all sources on the Internet are trustworthy, which could cause backfire if adopted blindly.
    
    (3) \ul{\textit{Model Deployment and Maintenance:}} Similar to the previous phase, downstream developers mentioned
    varying practices, though fewer compared to the previous stage were reflected in interviews and the survey (see Table~\ref{tab:thematic-results} and Figure~\ref{fig:RQ2-survey}), such as \textit{monitor input and output} and \textit{collect users' feedback}. This phase is particularly challenging due to limited tool support, unpredictable user interactions, and performance drift. Given the frequent failures for misuse and exploitation in the AI incident database, more frequent and careful practices in this stage are needed. The knowledge transfer gap remains for developers, as AI failure monitoring measures are proposed~\citep{lee2025a}, catering for unseen scenarios. Although it may seem unlikely that practitioners routinely consult academic literature, recent work has shown that many PTM reusers actively engage with ML papers to guide their development decisions~\citep{jiang2023empirical}. Thus, encouraging access to curated research insights—whether through papers or aggregator platforms—remains a promising direction for supporting safer deployment practices and achieving more robust AI safety.

\paragraph{Developers of PTMs.} As shown in Table~\ref{tab:thematic-results},  poor documentation is one of the biggest challenges for downstream developers, despite PTM developers' responsibility to provide clear information regarding potential failures and overall AI safety information. Empirical studies on current model cards efforts further highlight issues such as lack of detail and missing information~\citep{gao2024documenting}. To address this, PTM developers should use documentation templates such as model cards~\citep{mitchell2019model} as a good starting point, detailing pre-training data sources, model architecture and training details, clear evaluation results, intended use cases, and mitigation strategies, among others. Moreover, the evolution of open-source projects necessitates continuous documentation updates to ensure clarity and consistency for downstream users~\citep{gao2023add}. As PTM versions evolve (\eg architectural updates, expanded training data), documentation must be promptly updated to maintain clarity for users and support their ongoing efforts to achieve downstream AI safety.

\paragraph{Policy makers.} As observed in our results (Section~\ref{sec:RQ3_challenges}), downstream developers frequently reported a risk management policy interpretation gap, with 79\% of survey respondents agreeing that vague or absent guidelines hinder their mitigation efforts. For instance, guidelines like ``\textit{do not use untrusted third-party libraries}'' remain open to interpretation, while many stages simply lack actionable directives. The absence of concrete guidelines on addressing specific AI failures during each stage leads to varied approaches and confuses downstream developers. Smaller companies and open-source projects face even fewer regulations. For instance, P6, from a large company, must encrypt datasets before transfer, as mandated by company policies, whereas P2, from a small team, relies on a verbal agreement to prevent devices with sensitive data from leaving the office, without formal enforcement. This contrast underscores how organisational structures and policies influence developers' approaches to data concerns.

Despite efforts to apply AI safety frameworks in regulatory practices~\citep{nistRiskManagement, europaFuturiumEuropean}
and curate question banks for responsible AI risk assessment~\citep{lee2024qb4aira}, these initiatives emphasise principles over concrete development actions at each stage. While principles play a crucial role when dealing with abstract and complex parts of the engineering work, their effective implementation often requires translation into concrete actions. In many engineering fields, including software engineering, this can translate to structured artefacts such as checklist~\citep{garousi2016and,ngadiman2022checklist}.
We acknowledge the perspective of the traditional safety engineering community, which rightly cautions that checklist-based approaches are insufficient for guaranteeing robust system safety on their own, as they cannot replace rigorous hazard analysis or formal safety cases~\citep{leveson2004new, kelly1999arguing}. However, for general-purpose AI development, particularly in small companies and open-source projects where formal safety norms are entirely absent and resources for rigorous safety engineering are scarce, an evolving checklist outlining specific actions for AI failure mitigation serves as a transitional tool. It bridges the gap between high-level principles and the current void of practical engineering practices, helping to standardise initial mitigation efforts~\citep{braz2022less}. Some AI ethics checklists, such as the Deon Ethics Checklist,\footnote{See \url{https://deon.drivendata.org/data-science-ethics-checklist}.} have been investigated in previous studies~\citep{wong2023seeing}, but were found not to be easily accessible or practically usable by downstream developers, and there is still no concrete checklist for practical AI failure mitigation. Incorporating the idea of an AI failure mitigation checklist, along with the aforementioned information aggregator on useful strategies, would be helpful for downstream developers to navigate potential AI failures during PTM-reuse software development.

A frequent issue observed in the AI incident database as detailed in Section~\ref{sec:concerns} is that many AI-based software applications are used outside their intended purposes, leading to harmful consequences such as employees being fired solely based on AI decisions or non-medical certified software used for medical purposes.  \textcolor{blue}{In traditional software engineering, structured assurance 
frameworks such as Safety Integrity Levels (SILs) defined in 
IEC 61508~\citep{iec61508} and Evaluation Assurance Levels 
(EALs) defined in ISO/IEC 15408~\citep{EAL} provide tiered 
confidence levels for verifying that systems meet specific 
safety and security requirements --- ranging from basic 
functional testing to formally verified design.} Drawing an 
analogy to these established frameworks,  policy bodies could implement an automated assessment framework to evaluate AI systems across different safety dimensions or based on the aforementioned checklist. Indeed, emerging industry efforts such as the UL 3115~\cite{ul3115} standard are already beginning to provide such horizontal, cross-domain safety evaluation frameworks specifically for AI products. AI-based software meeting necessary safety requirements could receive tiered certification indicating its compliance level with specific failure mitigation standards, comparable to ISO 42001 certification~\citep{isoISOIEC420012023} but lightweight, tiered, and automated. This would help downstream users adopt AI-based software that meet established standards, ultimately fostering a higher degree of systemic AI safety. Stronger regulations are also needed to prevent misuse. 

In terms of the poor documentation issues reported by the developers in Section~\ref{sec:RQ3_challenges}, recent works on artefacts such as Software Bills of Materials (SBOMs) and SBOMs for AI software (AI-BOMs)~\citep{xia2023empirical, radanliev2024capability} highlights how supply-chain transparency can be strengthened across the AI ecosystem. While our study centres on downstream developers’ practices during PTM-reuse, these standards provide complementary mechanisms that policy makers could adopt to operationalise and audit the failure mitigation practices we identify. Full transparency from proprietary PTM providers remains unlikely in the near term, but requiring SBOM/AI-BOM style disclosures for downstream AI-based applications is more feasible and could substantially enhance accountability and user trust, bridging the gap between local engineering practices and global AI safety goals.

\paragraph{Researchers.} For software engineering researchers, our findings point to a knowledge transfer gap between research, policy and real-world implementation regarding approaching AI failures. In particular, 86\% of the survey participants agree that knowledge gap constitute a challenge for them to incorporate failure safeguards during the development (see Figure~\ref{fig:RQ3-survey}). Moreover, failure mitigation policy processing gap is consistently identified as a challenge across all the development stages (see Figure~\ref{fig:challenge-process}). Our analysis of the AI incident database reveals model reliability failures constitute more than half of the reported cases, while failures caused by irresponsible, unintended use or over-reliance of the delivered software is also observed taking 30.9\% of the incidents. Furthermore, 15.1\% of incidents involve both types of failures, highlighting the widespread premature adoption of unreliable AI-based software under unintended scenarios (see Table~\ref{tab:thematic-results}). Meanwhile, numerous research efforts have proposed solutions for specific technical AI failures~\citep{salewski2023incontext, chen2023holistic},  alongside evolving regulations on general AI safety~\citep{isoISOIEC420012023, europaFuturiumEuropean}. Some failures could have been prevented if relevant research were applied. For example, the over-reliance on AI without human oversight, such as cases involving employee layoffs solely based on AI decisions, could have been avoided or at least mitigated by adhering to human-in-the-loop requirements outlined in ISO 42001~\citep{isoISOIEC420012023}. However, with the rapid emergence of new research techniques and policy adjustments, downstream developers often struggle to keep pace, leading to information overload. The frequent occurrence of model behaviour failures and irresponsible usage, coupled with the abundant research and regulations, underscores a significant knowledge transfer gap between the general AI safety research and practical implementation of failure mitigation.

Therefore, research aimed at understanding the knowledge transfer gap between upstream AI researchers and regulators and downstream developers, and dissemination channels that can help bridge this divide would be valuable. Challenges related to knowledge dissemination should be investigated, as seen in prior studies on software engineering education that highlight difficulties in delivering relevant information to developers~\citep{fernandes2022devops}. Similarly, just as prior work has explored how failure stories can guide novice developers~\citep{anandayuvaraj2023incorporating}, investigating what types of failure-related information are most helpful for downstream developers could inform the design of practical tools and interventions for achieving overall AI safety. Regarding automated tool supports, developing recommender systems or summarisation techniques that process up-to-date policy documentation and extensive risk databases, such as the MIT AI Risk Repository~\citep{mitAIRisk}, could be a direction for future work. For example, TaskNavigator, proposed by \citet{treude2014extracting}, helps developers navigate software documentation based on different programming tasks. Similarly, such systems could automatically extract and filter relevant risk items from comprehensive taxonomies based on users' queries about specific AI failures at different stages of development. This approach could help standardise failure mitigation practices and alleviate information overload for downstream developers to improve the overall safety of AI-based software.

For researchers, the practices and challenges we identified also have implications in the context of AI agentic systems. AI agentic systems that decompose complex tasks into smaller subgoals~\citep{khot2022decomposed} and methodically explore different solution paths~\citep{yao2023tree} have attracted increasing attention as large language models advance. Although our study did not directly examine the development of agentic systems, we suggest that many of the challenges we identified are relevant to the development of such systems. For example, each subtask relying on a single model may face issues such as poor documentation, difficulty in monitoring and maintenance, and a lack of detailed, well-suited benchmark evaluations. However, these challenges could be amplified when multiple components are combined. For instance, iterative improvements to one model could inadvertently impact the behaviour of another, raising additional concerns about the interoperability and coordination of subtasks. This boundary erosion and entanglement of ML models are identified as hidden technical debt in machine learning systems~\citep{sculley2015hidden}. Additionally, Model Context Protocol (MCP)~\citep{hou2025model} emerged as a way for LLMs to interact with external tool. The engineering practices in the interface design and carried information, , such as the provision of explicit agent guidance files or skills configurations, will impact the overall reliability and safety of the agentic AI systems. Therefore, unique practices and challenges specific to agentic systems should be investigated in future studies to ensure comprehensive AI safety.

\section{Threats to Validity}

\paragraph{Construct validity:} Construct validity refers to how well a test or assessment measures the specific concept or characteristic it is designed to evaluate. The first construct in our study is the concept of AI failures, which we define as instances where the AI component in a system produces unintended, incorrect, or unsafe behaviour resulting in immediate technical harms. To align with prevailing vocabulary, our interview instrument used the term ``AI safety'' to prompt discussions. During the interviews, participants organically focused the discussion on concrete engineering failures rather than abstract concepts. Because our empirical data fundamentally centred on these immediate technical harms, broader implications involving indirect harms and existential risks, such as carbon emissions or super-intelligence, fell outside the scope of our study. Furthermore, because all subsequent survey questions and AI Incident Database annotations were derived directly from these empirically grounded interview themes, the construct validity was maintained across all three data sources.

The second construct is our division of the AI-based software development process involving PTM reuse into three stages: (1) preparation and PTM selection, (2) model development and testing, and (3) model deployment and maintenance. This structure is derived from several development processes of AI-based software~\citep{amershi2019software,jiang2023empirical}, and refined based on pilot study feedbacks to balance response quality and interview duration. While this three-stage framework provides a practical structure, finer granularity is possible. For example, the model development and testing stage could be further divided into data preparation, model training and adjustment, and model evaluation. Conducting interviews at this more detailed level could potentially yield more fine-grained insights.

Finally, we note that the SDLC-based figures (Figure~\ref{fig:practice-process} and Figure~\ref{fig:challenge-process}) is used primarily for illustrative purposes. It was derived from participants’ accounts to assist readers in contextualising how the identified practices and challenges relate to different development stages, rather than to propose a fully validated or prescriptive process model.

\paragraph{Internal validity}: Internal validity refers to the extent to which the observed results represent the truth in the population studied, and not due to methodological errors. For the survey participants, we have to rely on their self-claimed identity. We mitigate the risk by following the process by~\citet{schmidt2023accountability} to filter out participants whose answer in their developed AI-based software and job roles mismatch to ensure the quality of our survey responses. Another threat lies in the subjectivity of coding and synthesising process. To mitigate this, we conducted a rigorous examination of the transcripts, with three authors reviewing the coding assignments and aggregations. The final themes and subthemes were established through group consensus. Additionally, since interview participants did not necessarily develop the AI systems in the incident database, we did not establish direct links but provided directional discussions.

Additional threats to validity lie in the three data sources. We recruited five of the 16 interview participants from our professional networks, which could introduce selection bias. The downstream developers’ practices are self-reported rather than directly observed in real-world settings. However, the interview sessions were structured by development stage, with questions about AI failure concerns asked first and questions about practices asked afterwards, to help ensure that self-reported practices reflected actual experiences as closely as possible. There are also differences in demographics and system criticality across the data sources. In general, interview participants had more experience with deep learning and similar years of experience in software engineering compared to survey participants, while the AI incident database contained a higher proportion of safety-critical systems. Finally, the cases in the AI incident database were collected and reported by domain experts, which may not be comprehensive and could be subject to selection bias~\citep{anandayuvaraj2024fail}.

\paragraph{External validity}: External validity refers to the extent to which the findings of a study can be generalised to other settings. The first potential limitation is the relatively small number of interview participants. However, we reached subtheme-level saturation, and no new subtheme emerged with five additional participants. Second, we recruited participants from Hugging Face in descending order of organisation size, which may over-represent large companies. However, the participants' current roles involve diverse organisation sizes as indicated by the demographics. Third, we did not filter interview candidates based on the type of models or tasks they develop. As indicated by participants' demographics in Table~\ref{tab:demographics}, our results 
lean more towards NLP-based software developers. Fourth, our study primarily interviewed participants with experience in developing AI-based software that reuses CV and NLP components, and we do not claim generalisability beyond these two techniques.
Fifth, given the complexity of the development of AI-based software, each participant may not have experience in in all development stages.
To mitigate this, we continued recruitment until saturation was reached across all subthemes. Finally, our study primarily focuses on the development of general-purpose AI software. Consequently, our findings might not generalise to safety-critical systems, where safety regulations and risk mitigation practices are typically much more stringent.

\section{Conclusions}

We investigated the concerns, practices, and challenges downstream developers face regarding AI failures in the development of AI-based software. Our mixed-method approach triangulated qualitative interviews (N=16), a quantitative survey (N=86), and an analysis of 874 real-world cases from an AI incident database. The AI failure concerns reported by the downstream developers focus on data, model behaviour, and misuse/exploitation concerns. To address these concerns during development, downstream developers adopt several practices: (1) deeper understanding and assessment for failure risks, (2) regular monitoring and assessment for potential failures, (3) implementing technical safeguards, (4) maintaining documentation and expert consultation, and (5) relying on external parties and resources. The challenges they encounter include (1) infrastructure gaps, (2) technical and model-related difficulties, (3) lack of mature process and methodology, (4) resource constraints, (5) barriers to technical understanding and information access.

Our findings reveal that despite high awareness of potential AI failures, developers take limited action during the initial preparation and PTM selection stages. Additionally, knowledge gaps, missing policies, and a lack of tool support contribute to widely varying approaches to failure mitigation. The frequent occurrence of AI incidents, particularly related to model reliability, harmful and biased content, and misuse and exploitation, raises concerns about the effectiveness of these approaches. We provide implications for different stakeholders, emphasising the need to bridge the knowledge transfer gap between research/policy and downstream developers. 

\section{Data Availability} \label{sec:replication}
Due to data privacy constraints, we cannot share raw interview transcripts. However, the interview protocol, coding process, survey screenshot, survey responses, and AI Incident Database analysis are available at \url{https://github.com/Haoyu-Gao/AI-Safety-Mix-Study}.

\section{Declaration}

\textbf{Funding}: This project is not supported by external funding.

\noindent \textbf{Ethical Approval}: This project is approved by the Ethics Committee at the University of Melbourne under ID 30642.

\noindent \textbf{Informed Consent}: All participants provided informed consent before taking part in this research.

\noindent \textbf{Author Contributions}: Haoyu Gao contributed to idea formalisation, data collection, data analysis, and paper writing. Mansooreh Zahedi and Christoph Treude contributed to idea formalisation, data analysis, and paper writing. Wenxin Jiang and Hong Yi Lin contributed to data analysis. James C. Davis contributed to idea formalisation and paper writing.

\noindent \textbf{Data Availability Statement}: As mentioned in the above section, our data is available at \url{https://github.com/Haoyu-Gao/AI-Safety-Mix-Study}. However, we do not release the raw transcripts due to the ethical mandates.

\noindent \textbf{Conflict of Interest}: Not applicable.

\bibliographystyle{plainnat} 
\bibliography{reference}

\end{document}